\documentclass[ALICE,manyauthors]{cernphprep}
\usepackage[comma,square,numbers,sort&compress]{natbib}
\usepackage{hyperref}
\usepackage[T1]{fontenc}
\usepackage{xspace}
\usepackage[utf8]{inputenc}
\usepackage{booktabs}
\usepackage{multirow}
\usepackage{color}
\usepackage{euscript}
\usepackage{lineno}
\usepackage{orcidlink}

\begin{document}
%

\mathchardef\mhyphen="2D

\newcommand{\pp}           {pp\xspace}
\newcommand{\ppbar}        {\mbox{$\mathrm {p\overline{p}}$}\xspace}
\newcommand{\XeXe}         {\mbox{Xe--Xe}\xspace}
\newcommand{\PbPb}         {\mbox{Pb--Pb}\xspace}
\newcommand{\pA}           {\mbox{pA}\xspace}
\newcommand{\pPb}          {\mbox{p--Pb}\xspace}
\newcommand{\AuAu}         {\mbox{Au--Au}\xspace}
\newcommand{\dAu}          {\mbox{d--Au}\xspace}

\newcommand{\sigmapid}{$\sigma^{^{3}\mathrm{He}}_{\mathrm{d}E/\mathrm{d}x}$}
\newcommand{\nsigma} {$\left(  \mathrm{d}E/\mathrm{d}x -  \langle \mathrm{d}E/\mathrm{d}x \rangle_{^{3}\mathrm{He}} \right) /\sigma^{^{3}\mathrm{He}}_{\mathrm{d}E/\mathrm{d}x}$ }
\newcommand{\s}            {\ensuremath{\sqrt{s}}\xspace}
\newcommand{\snn}          {\ensuremath{\sqrt{s_{\mathrm{NN}}}}\xspace}
\newcommand{\pt}           {\ensuremath{p_{\rm T}}\xspace}
\newcommand{\meanpt}       {$\langle p_{\mathrm{T}}\rangle$\xspace}
\newcommand{\ycms}         {\ensuremath{y_{\rm CMS}}\xspace}
\newcommand{\ylab}         {\ensuremath{y_{\rm lab}}\xspace}
\newcommand{\etarange}[1]  {\mbox{$\left | \eta \right |~<~#1$}}
\newcommand{\yrange}[1]    {\mbox{$\left | y \right |~<~#1$}}
\newcommand{\dndy}         {\ensuremath{\mathrm{d}N_\mathrm{ch}/\mathrm{d}y}\xspace}
\newcommand{\dndeta}       {\ensuremath{\mathrm{d}N_\mathrm{ch}/\mathrm{d}\eta}\xspace}
\newcommand{\avdndeta}     {\ensuremath{\langle\dndeta\rangle}\xspace}
\newcommand{\dNdy}         {\ensuremath{\mathrm{d}N_\mathrm{ch}/\mathrm{d}y}\xspace}
\newcommand{\Npart}        {\ensuremath{N_\mathrm{part}}\xspace}
\newcommand{\Ncoll}        {\ensuremath{N_\mathrm{coll}}\xspace}
\newcommand{\dEdx}         {\ensuremath{\textrm{d}E/\textrm{d}x}\xspace}
\newcommand{\RpPb}         {\ensuremath{R_{\rm pPb}}\xspace}

\newcommand{\nineH}        {$\sqrt{s}~=~0.9$~Te\kern-.1emV\xspace}
\newcommand{\seven}        {$\sqrt{s}~=~7$~Te\kern-.1emV\xspace}
\newcommand{\twoH}         {$\sqrt{s}~=~0.2$~Te\kern-.1emV\xspace}
\newcommand{\twosevensix}  {$\sqrt{s}~=~2.76$~Te\kern-.1emV\xspace}
\newcommand{\five}         {$\sqrt{s}~=~5.02$~Te\kern-.1emV\xspace}
\newcommand{\twosevensixnn}{$\sqrt{s_{\mathrm{NN}}}~=~2.76$~Te\kern-.1emV\xspace}
\newcommand{\fivenn}       {$\sqrt{s_{\mathrm{NN}}}~=~5.02$~Te\kern-.1emV\xspace}
\newcommand{\LT}           {L{\'e}vy-Tsallis\xspace}
\newcommand{\GeVc}         {Ge\kern-.1emV/$c$\xspace}
\newcommand{\MeVc}         {Me\kern-.1emV/$c$\xspace}
\newcommand{\TeV}          {Te\kern-.1emV\xspace}
\newcommand{\GeV}          {Ge\kern-.1emV\xspace}
\newcommand{\MeV}          {Me\kern-.1emV\xspace}
\newcommand{\GeVmass}      {Ge\kern-.2emV/$c^2$\xspace}
\newcommand{\MeVmass}      {Me\kern-.2emV/$c^2$\xspace}
\newcommand{\lumi}         {\ensuremath{\mathcal{L}}\xspace}

\newcommand{\ITS}          {\rm{ITS}\xspace}
\newcommand{\TOF}          {\rm{TOF}\xspace}
\newcommand{\ZDC}          {\rm{ZDC}\xspace}
\newcommand{\ZDCs}         {\rm{ZDCs}\xspace}
\newcommand{\ZNA}          {\rm{ZNA}\xspace}
\newcommand{\ZNC}          {\rm{ZNC}\xspace}
\newcommand{\SPD}          {\rm{SPD}\xspace}
\newcommand{\SDD}          {\rm{SDD}\xspace}
\newcommand{\SSD}          {\rm{SSD}\xspace}
\newcommand{\TPC}          {\rm{TPC}\xspace}
\newcommand{\TRD}          {\rm{TRD}\xspace}
\newcommand{\VZERO}        {\rm{V0}\xspace}
\newcommand{\VZEROA}       {\rm{V0A}\xspace}
\newcommand{\VZEROC}       {\rm{V0C}\xspace}
\newcommand{\Vdecay} 	   {\ensuremath{V^{0}}\xspace}

\newcommand{\ee}           {\ensuremath{e^{+}e^{-}}} 
\newcommand{\pip}          {\ensuremath{\pi^{+}}\xspace}
\newcommand{\pim}          {\ensuremath{\pi^{-}}\xspace}
\newcommand{\kap}          {\ensuremath{\rm{K}^{+}}\xspace}
\newcommand{\kam}          {\ensuremath{\rm{K}^{-}}\xspace}
\newcommand{\pbar}         {\ensuremath{\rm\overline{p}}\xspace}
\newcommand{\kzero}        {\ensuremath{{\rm K}^{0}_{\rm{S}}}\xspace}
\newcommand{\lmb}          {\ensuremath{\Lambda}\xspace}
\newcommand{\almb}         {\ensuremath{\overline{\Lambda}}\xspace}
\newcommand{\Om}           {\ensuremath{\Omega^-}\xspace}
\newcommand{\Mo}           {\ensuremath{\overline{\Omega}^+}\xspace}
\newcommand{\X}            {\ensuremath{\Xi^-}\xspace}
\newcommand{\Ix}           {\ensuremath{\overline{\Xi}^+}\xspace}
\newcommand{\Xis}          {\ensuremath{\Xi^{\pm}}\xspace}
\newcommand{\Oms}          {\ensuremath{\Omega^{\pm}}\xspace}
\newcommand{\degree}       {\ensuremath{^{\rm o}}\xspace}

\begin{titlepage}
\PHyear{2022}       
\PHnumber{275}      
\PHdate{30 November}  

\title{Measurement of the production of (anti)nuclei in \\ \pPb collisions at \snn = 8.16 TeV}
\ShortTitle{Production of (anti)nuclei in \pPb collisions at \snn = 8.16 TeV}   

\Collaboration{ALICE Collaboration\thanks{See Appendix~\ref{app:collab} for the list of collaboration members}}
\ShortAuthor{ALICE Collaboration} 

\begin{abstract}

Measurements of (anti)proton, (anti)deuteron, and (anti)$^3$He production in the rapidity range \linebreak \mbox{$-1<y< 0$} as a function of the transverse momentum and event multiplicity in p--Pb collisions at a center-of-mass energy per nucleon--nucleon pair \snn = 8.16 TeV are presented. The coalescence parameters $B_2$ and $B_3$, measured as a function of the transverse momentum per nucleon and of the mean charged-particle multiplicity density, confirm a smooth evolution from low to high multiplicity across different collision systems and energies. The ratios between (anti)deuteron and (anti)$^3$He yields and those of \mbox{(anti)protons} are also reported as a function of the mean charged-particle multiplicity density. A comparison with the predictions of the statistical hadronization and coalescence models for different collision systems and center-of-mass energies favors the coalescence description for the deuteron-to-proton yield ratio with respect to the canonical statistical model.

\end{abstract}
\end{titlepage}

\setcounter{page}{2}

\section{Introduction} 
\label{sec:Introduction}
In ultra-relativistic hadronic collisions, light nuclei and antinuclei are produced in addition to other particles, but in very rare amounts with respect to the production of other light particles such as pions, kaons and protons.
The reduction of the yield of light (anti)nuclei for each additional nucleon measured at the  Large Hadron Collider (LHC) energies is about 300 in Pb--Pb~\cite{Acharya:2017bso} and about 1000 in pp collisions~\cite{Acharya:2017fvb}. 
Detailed measurements of the production of light (anti)nuclei, up to $^4$He, in the LHC energy regime have been carried out in recent years by the ALICE Collaboration~\cite{Adam:2015pna, Adam:2015vda, helium3_flow_5TeV, Acharya:2017fvb, Acharya:2017bso, deuteron_pp7TeV, Acharya:2019rys, Acharya:2019xmu, Acharya:2020sfy, Acharya:2020lus, Acharya:2020ogl, nuclei_pp_5TeV, nuclei_pp_13TeV}. 
The production of light nuclei has been studied extensively also at lower collision energies, from the AGS~\cite{ Bennett:1998be, Ahle:1999in, Armstrong:2000gz, Armstrong:2000gd}, to SPS~\cite{ Ambrosini:1997bf} and RHIC~\cite{Adler:2001prl,Adler:2004uy, Arsene:2010px,Agakishiev:2011ib,Adamczyk:2016gfs,Adam:2019wnb}.

Although the separation energy of nucleons and the binding energy of light nuclei are much smaller than the system temperature in heavy-ion collisions, their production yields are reasonably described within the statistical hadronization model (SHM)~\cite{SHM5,SHM6,SHM4,SHM2,SHM1,SHM3,Vovchenko:2022xil,Sharma:2022poi}, leaving open the question of their formation and survival in the post-hadronization phase. 
A different approach, based on the coalescence of protons and neutrons into a nucleus with mass number $A$, has also been developed~\cite{Coalescence3,Coalescence1, SATO1981153, Csernai:1986qf, Coalescence2,Blum:2017qnn,iEBE_VISHNU}. The coalescence probability is given by the parameter $B_A$, defined as
\begin{equation}
B_{A} = { \biggl( \dfrac{1}{2 \pi p^{A}_{\mathrm T}} \dfrac{ \mathrm{d}^2N_A}{\mathrm{d}y\mathrm{d} p_{\mathrm T}^{A}}  \biggr)}  \bigg/{  \biggl( \dfrac{1}{2 \pi p^{\rm p}_{\mathrm T}} \dfrac{\mathrm{d}^2N_{\mathrm p} }{\mathrm{d}y\mathrm{d}p_{\mathrm T}^{\mathrm p}} \biggr)^A}  ,
\label{eq:BA}
\end{equation}
where the labels $A$ and p indicate the nucleus with mass number $A$ and the proton, respectively, \pt is the transverse momentum and \mbox{$p_{\mathrm T}^{\rm p}$ = $p_{\mathrm T}^{A}$/$A$}. In such a model, neutrons and protons are assumed to have the same production spectra, since both belong to the same isospin doublet.

State-of-the-art SHM and coalescence model provide similar predictions for the yields of \mbox{(anti)nuclei}~\cite{BRAUNMUNZINGER2019144,CoalescenceTheory,Mrowczynski:2020ugu}. 
Possibilities to discriminate between the two approaches could come from the study of the production yields of different nuclei that differ in size. The coalescence model, indeed, is sensitive to the size of the nucleus, in particular to the relation between nuclear size and emission source size~\cite{SATO1981153,Coalescence2}. On the contrary, the predictions of the SHM depend only on the mass and on the spin degeneracy factor of the nucleus. 
In a simple coalescence model, in which the size of the emitting source is not taken into account, the $B_A$ parameter is expected to be independent of transverse momentum, multiplicity and source size. However, previous experimental results have shown that $B_A$ at a given \pt weakly depends on multiplicity in pp collisions~\cite{deuteron_pp7TeV,Acharya:2020sfy}, while in Pb--Pb collisions it shows a strong decrease with multiplicity~\cite{Anticic:2004yj,Adam:2015vda, Acharya:2019xmu}. 
From different femtoscopy measurements at different multiplicities~\cite{Abelev:2014pja}, it is known that the source radius ($R$) is related to the average charged particle multiplicity density ($\langle$d$N_{\mathrm{ch}}/$d$\eta_{\mathrm{lab}}\rangle$) through the following parameterization: 
\begin{equation}
 R =\rm a \langle \rm{d} \it{N}_{\mathrm{ch}}/ \rm{d}\eta_{\mathrm{\rm{lab}}}\rangle^{
1/3} + b ,
\end{equation}
where a and b are free parameters~\cite{CoalescenceTheory}. 
Therefore, the mean charged-particle multiplicity density allows the comparison of different collision systems at similar nuclear emission volumes. 
The p--Pb collision system covers multiplicities that are between pp and Pb--Pb collisions and offers the possibility to explore intermediate source sizes. 

The production mechanism of light (anti)nuclei can be further investigated by comparing the ratio of their yields to those of protons for different multiplicities with the SHM and coalescence model predictions. 

In a different context, the study of light (anti)nuclei production in high-energy pp and p--A collisions could provide new inputs to the understanding of the abundance of antimatter searched in space experiments~\cite{Donato:2008yx, Korsmeier:2017xzj, Doetinchem_2020}, such as AMS-02~\cite{Kounine:2012ega} and GAPS~\cite{Hailey:2009fpa}. The observation of antinuclei with a mass larger than that of antiprotons could be related either to segregated primordial antimatter or to the annihilation of dark matter particles in the galactic halo~\cite{Chang:2018bpt}. In this respect, the understanding of the production rate of antinuclei stemming from energetic collisions at LHC energies is an important input for estimates of the background originating from interactions between cosmic rays and the interstellar medium.

Previous articles from the ALICE experiment have already reported results on the production of (anti)protons and light (anti)nuclei in p--Pb collisions at a center-of-mass energy per nucleon--nucleon pair \snn = 5.02 TeV~\cite{ALICE:2013wgn, Acharya:2019xmu, Acharya:2019rys}. This article reports new results on the production of (anti)protons, (anti)deuterons, and (anti)$^3$He nuclei in p--Pb collisions at \snn = 8.16 TeV. Transverse momentum differential yields of (anti)protons in seven multiplicity classes, of (anti)deuterons in four multiplicity classes, and of (anti)$^3$He in the multiplicity-integrated data sample are presented. Antinuclei-to-nuclei ratios are found to be consistent with unity~\cite{Acharya:2017fvb,Adam:2015vda,Acharya:2019xmu,Acharya:2019rys,Acharya:2020sfy} as expected from coalescence and thermal models and the (anti)baryon symmetry at midrapidity at LHC energies~\cite{Baryon_ppcollisions}.
Coalescence parameters $B_2$ and $B_3$ are studied as a function of \pt/$A$. In addition, the values of $B_2$ as a function of the mean charged-particle multiplicity density are compared to the results obtained for different collision systems and energies, as well as to model predictions.

\section{Experimental apparatus and data sample} 
\label{sec:ExperimentalApparatus}

ALICE is a general-purpose detector system designed to study heavy-ion collisions at the LHC. Thanks to its excellent tracking and particle identification (PID) capabilities, ALICE is ideally suited to study light nuclei and antinuclei in different collision systems. A detailed description of the ALICE subsystems and their performance can be found in Refs.~\cite{ALICE_general_2, ALICEperformance} and references therein.

The detectors used in this analysis are the Inner Tracking System (\ITS), the Time Projection Chamber (\TPC) and the Time-Of-Flight (\TOF) detector. These detectors are located in the central barrel inside a solenoidal magnet with a field strength of {\it B} = 0.5 T.

The \ITS~\cite{ALICE_general_2,ITS}, which covers the full azimuthal angle and the pseudorapidity interval $|\eta_{\rm lab}|$ < 0.9, is mainly used to determine the primary vertex of the collision and the secondary vertices from weak decays for the track reconstruction. It is composed of three subsystems of silicon detectors placed around the nominal beam axis with a cylindrical symmetry: the two innermost layers are Silicon Pixel Detectors (\SPD), followed by two layers of Silicon Drift Detectors (\SDD), and finally by two layers of Silicon Strip Detectors (\SSD). As discussed in the next Section, the \ITS is also used to separate primary nuclei from secondary nuclei produced in the interaction of primary particles with the detector material, via the determination of the distance-of-closest approach (DCA) of the track to the primary vertex.

The \TPC~\cite{TPC} is the main tracking detector and allows for particle identification by measuring the specific ionization energy loss (\dEdx) in the gas. The \TPC is a cylindrical drift chamber, coaxial with the beam pipe and filled with a gas mixture containing 90\% Ar and 10\% CO$_2$ at atmospheric pressure. With a radius ranging from 85 to 250 cm and a length of 500 cm in the beam direction, the \TPC volume occupies the same pseudorapidity interval covered by the \ITS. The trajectory of a charged particle is estimated by measuring the gas ionization in up to 159 samples (clusters) along a path of about 160 cm. 
The \TPC provides a measurement of the charged-particle transverse momentum with a resolution ranging from about 1\% at 1 $\mathrm{GeV}/\textit{c}$ to about 3\% at 10 $\mathrm{GeV}/\textit{c}$.  By combining the \TPC tracking capabilities with the ones of the \ITS and Transition Radiation Detector (\TRD)~\cite{Acharya:2017lco}, the transverse momentum resolution improves by a factor 5--7 depending on the \pt~\cite{ALICE_general_2}. Moreover, the \TPC provides a measurement of the specific energy loss with a resolution ranging from about 5.2\% in pp collisions to about 6.5\% in central Pb--Pb collisions~\cite{ALICEperformance}, for minimum ionizing particles crossing the full detector.

The \TOF~\cite{TOF} detector covers the full azimuth and the pseudorapidity interval $|\eta_{\rm lab}|$ < 0.9. 
The detector is made of Multi-gap Resistive Plate Chambers (MRPC) located at an average distance of 380 cm from the beam axis. 
The TOF time resolution is 56 ps~\cite{Carnesecchi:2018oss}, while the event time resolution varies depending on the collision system and on the track multiplicity~\cite{Adam:2016ilk}. 
The best precision on the event time ($t_0$) evaluation is obtained by using the TOF detector itself, with a resolution better than 20 ps if more than 50 tracks are used for its determination, which is the case of high multiplicity p--Pb collisions. 
The start time for the time of flight is provided by the T0 detector~\cite{T0_Cortese:2004aa} and by the TOF detector itself, the latter being particularly useful for measurements at large multiplicities. 
The T0 consists of two arrays of Cherenkov counters, T0A and T0C, placed on opposite sides of the nominal interaction point, covering the pseudorapidity regions $4.6 < \eta_{\rm lab} < 4.9$ and $-3.3 < \eta_{\rm lab} < -3.0$. A weighted average is performed when both T0 and TOF detectors have measured the start time~\cite{Adam:2016ilk}. 
Particles are identified by comparing the measured time-of-flight with that evaluated from track momentum and length for each mass hypothesis.

The \VZERO detector~\cite{T0_Cortese:2004aa} consists of two plastic scintillator arrays (\VZEROA and \VZEROC) located at asymmetric positions, one on each side of the interaction point, and covering the pseudorapidity regions $2.8 < \eta_{\rm lab} < 5.1$ and $-3.7 < \eta_{\rm lab} < -1.7$. The V0 detector is used to define the minimum-bias trigger and to select events based on multiplicity. The (anti)proton and (anti)deuteron analyses were carried out also for several multiplicity classes, defined as percentiles of the V0 signal~\cite{VZEROPerformance}: 0--5\%, 5--10\%, 10--20\%, 20--40\%, 40--60\% , 60--80\%, and 80--100\% for the former and 0--10\%, 10--20\%,  20--40\%, and 40--100\% for the latter.

The analysis uses a sample of 40 million minimum-bias events collected by ALICE in 2016 during the LHC p--Pb campaign at \snn = 8.16 TeV. Two configurations of colliding beams were used, one corresponding to the proton beam traveling in the direction from \VZEROA to \VZEROC, while $^{208}$Pb ions circulated in the opposite direction (denoted by p--Pb), the other corresponding to a reversed direction of both beams (denoted by Pb--p). Due to the high interaction rate available in the 2016 data taking (about 100 kHz), a fraction of the triggered events contains data corresponding to more than one collision (pile-up). Events with multiple vertices identified with the SPD are tagged as pile-up and removed from the analysis. The amount of pile-up events is about 2$\%$ of the total~\cite{Acharya:2020puh}. The selected events are those in which the colliding ions interact via inelastic collisions and at least one charged-particle is produced in the central pseudorapidity region $|\eta_{\rm lab}|$ < 1. This event class is referred to as \mbox{INEL > 0}.

Finally, only events with a reconstructed primary vertex position along the beam axis within 10 cm from the nominal interaction point are selected. In this analysis, the production of primary (anti)nuclei is measured in a rapidity window --1 < $y$ < 0 in the center-of-mass system. Since the energy per nucleon of the proton beam is higher than that of the Pb beam, the nucleon--nucleon center-of-mass system is shifted with respect to the laboratory frame by 0.465 units of rapidity in the direction of the proton beam.

\section{Analysis procedure} 
\label{sec:Analysis}
In this section, the analysis procedure is described. Specifically, the criteria used for the track selection, the signal extraction, the corrections based on Monte Carlo (MC) simulations, and the evaluation of the systematic uncertainties are detailed and discussed. 

\subsection{Track selection and particle identification} 
\label{subsec:TrackSelection_PID}

(Anti)proton, (anti)deuteron, and (anti)$^{3}$He candidates are selected from a sample of charged-particle tracks reconstructed in the ITS and TPC in the pseudorapidity range $|\eta_{\rm lab}| < 0.8$. Several track quality criteria are applied, such as a minimum number of clusters in the TPC of at least 70 out of a maximum of 159, and in the ITS of at least 2 with one cluster located in any of the two innermost ITS layers. For (anti)protons, at least three clusters in the SDD and SSD are also requested. With such a small number of tracking points, the probability of having tracks with wrongly associated clusters is not negligible. This contribution is strongly suppressed by applying a selection on the quality of the track reconstruction procedure, namely $\chi^{2}$ per ITS cluster $<2.5$.
A good quality of the track fit is also needed, hence the $\chi^{2}$ per TPC reconstructed point is required to be less than 4.
The number of TPC clusters used in the \dEdx calculation is required to be larger than 50 to ensure a good \dEdx resolution.
The contribution from secondary tracks is reduced by requiring a maximum DCA to the primary vertex in the transverse plane ($\mathrm{DCA}_{xy}$) and in the longitudinal direction ($\mathrm{DCA}_{z}$) lower than 0.1 cm for (anti)deuterons and (anti)$^3$He, and lower than 0.0105 + 0.0350/\pt$^{1.1}$ cm and 2 cm, respectively, for (anti)protons.  

(Anti)protons are identified by exploiting the combined information from the specific energy loss measured in the ITS and in the TPC, in the transverse momentum range between 0.3 and 0.8 GeV/$c$, and the information from the TPC and TOF in the \pt range between 0.8 and 2.5 GeV/$c$. For the ITS analysis, the value of \dEdx measured by the four outer layers of the ITS (two layers of SDD and two layers of SSD) is used. As the energy loss distribution in the silicon layers of the ITS is a Landau distribution with a typical long tail, a truncated mean approach is chosen in order to reduce the energy loss fluctuations~\cite{ITS}. The particle identification procedure consists in assigning a track to a particle species depending on the distance from the expected Bethe--Bloch parameterization. 
The (anti)protons are identified using the specific energy loss inside the active volume of the ITS, and (anti)nuclei using that of the TPC. For this purpose, the distribution of n$\sigma^{\rm ITS, TPC}$ = (\dEdx $-$ $\langle \mathrm{d}E/\mathrm{d}x \rangle$)/$\sigma$ is extracted, being $\langle \mathrm{d}E/\mathrm{d}x \rangle$ the expected average \dEdx for the corresponding (anti)particle, and $\sigma$ the ITS or TPC \dEdx measured resolution. 
For the (anti)$^{3}$He identification, the \dEdx measured in the TPC is required to be within 3$\sigma$ from the expected average for (anti)$^{3}$He in the full \pt range covered by these measurements (1.5 $\mathrm{GeV}/\textit{c}$ $< p_{\mathrm T}<$ 6.0 $\mathrm{GeV}/\textit{c}$). 

(Anti)deuterons with \pt  $< 1.2\ \mathrm{GeV}/\textit{c}$ and (anti)protons with 0.8$<\pt<$2.5 GeV/$c$ are identified by requiring that the measured energy loss is within 3$\sigma$ from the corresponding expected average. 
For (anti)deuterons with \pt  $> 1.2\ \mathrm{GeV}/\textit{c}$ and (anti)protons with \pt  $> 0.8\ \mathrm{GeV}/\textit{c}$, the information from the TOF detector is used in addition and the signal is extracted from a fit to the n$\sigma^{\rm TOF}$ = $(\Delta t - \Delta t_{\mathrm{d}}) / \sigma_{\mathrm{TOF}}$ distribution, where $\Delta t$ is the measured time-of-flight, $\Delta t_{\mathrm{d}}$ its expected value for protons or deuterons and $\sigma_{\mathrm{TOF}}$ the resolution on the time-of-flight measurement. The fit function consists of a Gaussian with an exponential tail for the signal and the sum of two exponential functions for the background. 
The signal is extracted by integrating the signal function in an asymmetric range centered at $\mu_{0}$ (mean of the Gaussian signal), which is slightly different from zero because of small miscalibration effects of the TOF detector: [$\mu_{0} -3\sigma_{\mathrm{TOF}}$, $\mu_{0} +3.5\sigma_{\mathrm{TOF}}$]. 
A similar shift in the peak position of the TOF signal was also observed in~\cite{Adam:2015pna}.
The background in the TOF response is negligible for (anti)deuterons with \pt $\leq$ 1.6 $\mathrm{GeV}/\textit{c}$ and increases up to 60\% going to high \pt.

\subsection{Corrections} 
\label{subsec:Corrections}
The raw spectra are corrected using a MC simulation taking into account the conditions of the detector during the data acquisition. Particles are generated with a uniform distribution in transverse momentum and rapidity, within \mbox{0 < \pt < 10 $\mathrm{GeV}/\textit{c}$} and --1 < $y$ < 1, respectively. (Anti)nuclei are injected on top of the underlying event generated by HIJING~\cite{HIJING}. For particle propagation and simulation of the detector response, the GEANT4 package is used~\cite{GEANT}.

\subsubsection{Background from spallation and weak decays of (anti)\texorpdfstring{$^{3}_{\Lambda}$}{}\texorpdfstring{\rm{H}}{}} 
\label{subsec:secondary}
The interaction of primary particles and nuclei in the detector material produces nuclear fragments, called spallation products. The DCA distributions of primary and secondary nuclei are different. The tracks of primary nuclei point to the primary vertex and therefore have a narrow distribution peaked at zero. 
Antinuclei cannot originate from the detector material. 
Spallation fragments, instead, show a broader and flatter distribution. Therefore, secondary nuclei produced by spallation can be discriminated using the DCA of their reconstructed tracks to the primary vertex~\cite{Adam:2015vda}. The preselection on the DCA$_z$ (see Sect.~\ref{subsec:TrackSelection_PID}) reduces the background, without affecting primary nuclei. 

To remove the residual contribution of secondary nuclei, a template fit of the DCA$_{xy}$ distributions is used, as done in previous measurements~\cite{Acharya:2019rys, Acharya:2020sfy}. 
The templates describing secondary deuterons from material are obtained from MC simulations. 
The primary deuteron distribution is described by using antideuteron distributions from data, since antideuterons do not have contribution from secondary nuclei from material. The fit is performed in the range $|$DCA$_{xy}|<$ 0.9 cm. 
The contamination of secondary deuterons amounts to about 15$\%$ in the lowest \pt interval (0.6 < \pt < 0.8 $\mathrm{GeV}/\textit{c}$) and decreases exponentially towards higher \pt until it becomes negligible above 1.6 $\mathrm{GeV}/\textit{c}$. The limited number of $^3$He candidate tracks, instead, does not allow for a background subtraction based on this method. 
Therefore, the fraction of secondary $^3$He and the corresponding uncertainty are taken from the previous analysis of p--Pb collisions~\cite{Acharya:2019xmu}, under the assumption that the relative contribution of secondary $^3$He from spallation does not change significantly from \snn = 5.02 TeV to \snn = 8.16 TeV. This corresponds to a secondary fraction of approximately 27$\%$ in the transverse momentum interval 1.5 < \pt < 2 $\mathrm{GeV}/\textit{c}$ and is negligible for higher \pt.

Another background contribution is given by secondary (anti)$^3$He from mesonic weak decays of (anti)$^3_\Lambda $H~\cite{MesonicDecayHypertriton} (secondary nuclei from feed-down). 
The contribution of secondary (anti)$^3$He from feed-down is estimated using MC simulations and then subtracted from the inclusive $p_{\mathrm T}$ distribution, as described in Ref.~\cite{Acharya:2019xmu}. The fraction of secondary (anti)$^3$He from feed-down is given by:

\begin{equation}
f_{\text{feed-down}}(p_{\mathrm T}) = \dfrac{\epsilon_{\text{feed-down}}(p_{\mathrm T})}{\epsilon_{^3\mathrm{He}}(p_{\mathrm T})} \times BR \times \dfrac{ ^3 _\Lambda \mathrm{H} }{^3\rm He} .
\label{eq:feeddown}
\end{equation}

The (anti)$^3_{\Lambda}$H-to-(anti)$^3$He ratio is extrapolated to the integrated multiplicity class of p--Pb collisions at \snn = 8.16 TeV, using the measured ratio as a function of $\langle$d$N_{\mathrm{ch}}/$d$\eta_{\mathrm{lab}}\rangle$ in Pb--Pb collisions at \mbox{$\sqrt{s_{\mathrm {NN}}}$ = 2.76 TeV}~\cite{Hypertriton}, assuming a linear trend. 
As a cross-check, this ratio is compared to the expectations of the canonical statistical model~\cite{Vovchenko:2018fiy}, resulting in good agreement. 
The $BR$ represents the branching ratio of the mesonic decay of $^3 _\Lambda $H, which amounts to about 25$\%$, as reported in Ref.~\cite{MesonicDecayHypertriton}. 
Monte Carlo simulations are used to evaluate the fraction of (anti)$^3_\Lambda $H that passes the track selection in the $^3$He (anti)nucleus channel. This fraction is estimated using the ratio of the reconstruction efficiency of secondary (anti)$^3$He from (anti)$^3 _{\Lambda}$H decays ($\epsilon_{\text{feed-down}}$) to the reconstruction efficiency of primary (anti)$^3$He ($\epsilon_{^3\mathrm{He}}$). 
The fraction of secondary nuclei from feed-down is about 4$\%$ with a weak dependence on transverse momentum.

The relative contribution of secondary (anti)deuterons from (anti)$^3 _{\Lambda}$H is estimated with the same method and is found to be negligible, due to the much larger abundance of primary deuterons with respect to $^3 _{\Lambda}$H.

\subsubsection{Acceptance and efficiency correction} 
\label{subsec:efficiency}
The reconstruction Acceptance $\times$ Efficiency (\emph{A} $\times$ $\epsilon$) is defined as the ratio between the reconstructed primary tracks, in the rapidity and pseudorapidity regions of interest, and the generated particles in the same rapidity interval, as given by:

\begin{equation}
\emph{A} \, \times \, \epsilon = \dfrac{\it N_{\mathrm{rec}, |y|<0.5, |\eta_{\rm lab}|<0.8}}{\it N_{\mathrm{gen,} |y|<0.5}} .
\label{eq:efficiency}
\end{equation}

The same track selection criteria used for data are applied to the reconstructed tracks in MC, in order to select only (anti)nuclei of our interest, namely (anti)protons, (anti)deuterons and (anti)$^3\rm{He}$. 
The efficiency for the antinuclei is reduced compared to that of nuclei because of annihilation processes with the beam pipe and the detector material. The efficiency for (anti)deuterons depends on \pt and it ranges between 35 and 70$\%$ in the region of \pt < 1.2 $\mathrm{GeV}/\textit{c}$, where the analysis is performed using only the TPC information, while it is around 50$\%$ for \pt > 1.2 $\mathrm{GeV}/\textit{c}$, because of the additional requirement of having a hit in the TOF detector for (anti)deuteron tracks. The latter implies the crossing of the TRD and part of the support structure, which are located between the TPC and the TOF detector. The efficiency of (anti)protons ranges between 20 and 60$\%$ in the low \pt region (0.3$<\pt<$0.8 GeV/$c$) where the analysis is done using ITS and TPC, while it is $\sim$70$\%$ in the region where the analysis is done with TOF (0.8$<\pt<$2.5 GeV/$c$). Finally, the efficiency of (anti)$^3$He is $\sim$75$\%$ and only mildly dependent on \pt, since the analysis is performed using the TPC only.

\subsubsection{Signal and event loss} 

An additional correction is related to the event and signal loss due to the trigger efficiency. 
In order to account for the INEL > 0 events that are erroneously rejected (event loss) and for all the (anti)nuclei lost because they were produced in the wrongly rejected events (signal loss), MC simulations are used to correct the measured \pt spectra. The corrected spectrum is given by

\begin{equation}
\frac{1}{N^{\rm INEL>0}_{\rm events}} \frac{\mathrm{d}^2N_{\rm corr}}{\mathrm{d}y\mathrm{d}p_{\rm T}} = \frac{1}{\frac{N_{\rm events}}{\epsilon_{\rm event}}} \frac{1}{\epsilon_{\rm signal}}\frac{\emph{f}_{primary}}{\emph{A} \times \epsilon} \frac{\mathrm{d}^2N}{\mathrm{d}y\mathrm{d}p_{\rm T}} ,
\end{equation}

where $N^{\rm INEL > 0}_{\rm events}$ is the number of INEL > 0 events, $N_{\rm events}$ is the number of selected events which fulfill the event selection criteria, $\epsilon_{\rm event}$ is the event selection efficiency, $\epsilon_{\rm signal}$ is the (anti)nuclei reconstruction efficiency, $f_{\rm primary}$ is the primary fraction discussed in Sect.~\ref{subsec:secondary} and $A \times \epsilon$ is the efficiency estimated in Sect.~\ref{subsec:efficiency}.
The event selection and (anti)nuclei reconstruction efficiencies are estimated using MC simulations with the same procedure followed in Ref.~\cite{Acharya:2020sfy}. 
Both, $\epsilon_{\rm event}$ and $\epsilon_{\rm signal}$ are approximately 90$\%$ in all multiplicity classes, the latter independently of \pt.  

\subsection{Systematic uncertainties} 
\label{subsec:Systematics}

The different contributions to the systematic uncertainties of (anti)protons, (anti)deuterons and (anti)$^3$He are summarized in Table~\ref{tab:Systematics}. The total systematic uncertainty is calculated as the sum in quadrature of each contribution. 
The improvement of the systematic uncertainties of (anti)nuclei with respect to previous measurements~\cite{deuteron_pp7TeV} is mainly due to a better knowledge of the detector and therefore to its better implementation in the MC simulations. Moreover, the recent results on the absorption studies of (anti)deuterons~\cite{antideuteronInelCS} allowed for a better treatment of the systematic uncertainties related to the hadronic interaction.

\begin{table}
\centering
\caption{Summary of the different contributions to the overall systematic uncertainties, reported for (anti)protons, (anti)deuterons, and (anti)$^3$He in the lowermost and uppermost \pt intervals. All values are given in percentage. }
\begin{tabular}{ccccccc}
\toprule
  \rm{Particle } & \multicolumn{2}{c}{\rm{p ($\overline{\mathrm p}$)}} & \multicolumn{2}{c}{\rm{d ($\overline{\mathrm d}$)}} & \multicolumn{2}{c}{\rm{ $^3$He ($^3\overline{\mathrm{He}}$)}} \\ 
\midrule
  \pt range (GeV/$c$)& \rm{0.30 -- 0.35} & \rm{2.4 -- 2.5} & \rm{0.6 -- 0.8} & \rm{ 3.4 -- 4.2}  & \rm{1.5 -- 2.0} & \rm{3.0 -- 6.0} \\
\midrule
  \rm{Source of uncertainty} & & & & & &\\
\midrule
Tracking  	& 4.2 (4.0) & 6.0 (8.5) & 1.6 (2.3)  &  1.1 (2.5) 		& 2.6 (4.4) & 2.6 (4.4) \\
ITS--TPC matching & -- & 1.0 (1.0) & 1.0 (1.0)  &  1.0 (1.0) & 1.0 (1.0)  &  1.0 (1.0) \\
Signal extraction & 8.0 (8.5) & 0.5 (0.5) & 1.4 (1.3)  &  3.3 (3.0) & 	2.7 & negl. \\
Secondaries material & 1.0 (1.0) & 1.0 (1.0)  &  0.2 	& negl. &  9.0  & negl.   \\
Secondaries feed-down & negl. & negl. & -- & -- &   3.4 (3.5) & 2.6 (2.8)  \\
Material Budget	& 4.5 (6.0) & negl. &  1.0 (1.0)   & 1.0 (1.0) & 0.2 (0.3)     &  0.2 (0.3)  \\
Hadronic interaction & negl. & negl. &  0.6 (1.0)   & 1.5 (6.0) 	&  1.0 (3.0) & 1.0 (3.0)     \\
TPC-TOF matching & -- & 1.0 (1.0) & --  &  4.2 (5.2)  & -- & -- \\
Signal loss correction & 3.0 (3.0) & 3.0 (3.0)  &  1.0 (1.0) & 1.0 (1.0) & -- & -- \\
\midrule
Total & 10.5 (11.5) & 7.0 (9.2) &  3.6 (3.4)   & 5.9 (6.8) 	&  10.3 (6.4) & 3.8 (6.0)     \\
\bottomrule
\end{tabular}
\label{tab:Systematics}
\end{table}

For the tracking-related systematic uncertainty, the selection criteria used in the track selection discussed in \mbox{Sect.~\ref{subsec:TrackSelection_PID}} are varied, both in data and MC, using a random uniform distribution around the nominal value. The relative systematic uncertainty is given by the root mean square (RMS) divided by the mean value of the distributions of the (anti)deuteron and (anti)$^3$He corrected yields in each \pt interval. 
Variations consistent with statistical fluctuations are rejected from the trials used to estimate the systematic uncertainties using the Barlow criterion~\cite{Barlow:2002yb}. This contribution is between 4.0$\%$ and 8.5$\%$ for (anti)protons, between 1.1$\%$ and 2.5$\%$ for (anti)deuterons, and between 2.6$\%$ and 4.4$\%$ for (anti)$^{3}$He, depending on \pt.

The difference between the ITS--TPC matching efficiency in data and MC is accounted for as a relative systematic uncertainty contribution of 1$\%$ for all species.

To assess the systematic uncertainty due to the signal extraction of (anti)deuterons, the $n\sigma_{\mathrm dE/\mathrm dx}$ interval used for the selection of candidates as well as the signal extraction range are varied and the spread of the efficiency-corrected yield in each transverse momentum interval is considered as systematic uncertainty. The contribution coming from the difference between the bin-counting method and the integral of the signal function is added in quadrature to the previous one. The systematic uncertainty due to the signal extraction of (anti)deuterons is between 1.3$\%$ at low \pt and 3.3$\%$ at high \pt. 

The systematic uncertainty on the (anti)$^3$He signal extraction is given by the difference between the yields obtained by subtracting the $^3$H contribution using a Gaussian and an exponential function. This contribution is 2.7$\%$ and is relevant only in the transverse momentum interval 1.5 < \pt < 2.0 GeV/$c$.

The systematic uncertainty due to the estimate of secondary nuclei from spallation processes is obtained by varying the DCA$_{xy}$ and DCA$_z$ selection ranges as well as the bin width of the histograms and the fit range. 
For protons this contribution is around 1$\%$, while for deuterons it is at most 0.2$\%$ at low \pt and decreases exponentially becoming negligible for \pt > 1.4 $\mathrm{GeV}/\textit{c}$. For $^{3}$He, the systematic uncertainty due to spallation background is 9$\%$, taken from Ref.~\cite{Acharya:2019xmu} for the first \pt interval, and an additional 3$\%$ contribution is assigned to the second \pt interval as a conservative estimate based on the results obtained in Ref.~\cite{Acharya:2019xmu}. 

For (anti)$^3$He, the systematic uncertainty due to the feed-down correction is also estimated. This uncertainty is given by half of the difference between the maximum and the minimum values obtained by repeating the linear extrapolation of the $^3 _\Lambda $H-to-$^3$He ratio moving upwards and downwards the average values by their uncertainties. This contribution ranges between 2.6$\%$ and 3.5$\%$.  

The systematic uncertainty due to the limited precision of the description of the detector and support structure material is estimated as half of the difference between the efficiencies obtained by increasing and decreasing in MC simulations the ALICE material budget by 4.5$\%$. This value corresponds to the current uncertainty on the material obtained by photon conversion measurements~\cite{ALICEperformance}.
The resulting systematic uncertainty is, for protons (antiprotons), 4.5 \% (6.0 \%) at low \pt and negligible at high \pt, while it is approximately 1$\%$ for (anti)deuterons and at most 0.3$\%$ for (anti)$^3$He, independent of \pt.  

The uncertainty on the hadronic interaction cross section of (anti)deuterons with the detector materials results in an uncertainty on the measured (anti)deuteron \pt spectrum. This contribution is calculated using the existing experimental measurements of the (anti)deuteron inelastic cross sections on different targets~\cite{absorption1,absorption2,absorption3,absorption4}, and the first measurement for low-energy antideuterons performed by ALICE~\cite{antideuteronInelCS}. 
The available experimental data are fitted simultaneously using the Glauber model parameterizations of GEANT4. In this fit, the momentum dependencies of the GEANT4 cross sections for the different targets are fixed and the only free parameter is a scaling factor which is determined with its uncertainty.
The (anti)deuteron reconstruction efficiency is then calculated in the simulations by increasing and decreasing the scaling factor by the obtained uncertainty. 
The systematic uncertainty due to the hadronic interaction is given by half of the difference between these efficiencies. This contribution is approximately 0.6$\%$ (1$\%$) at low \pt and 1.5$\%$ (6$\%$) at higher \pt for deuterons (antideuterons).
A similar procedure gives 1$\%$ for $^3$He and 3$\%$ for anti$^3$He.

In the (anti)deuteron spectrum for \pt > 1.2 $\mathrm{GeV}/\textit{c}$, where the TOF is used for the signal extraction, the uncertainty on the material thickness of the TRD and part of the space frame that is located between the TPC and the TOF is also considered. The uncertainty on the material thickness results in an uncertainty on the TPC--TOF matching efficiency, which is given by $\epsilon=e^{- x/ \lambda^{\rm d}_{\rm I}}$, where $x$ is thickness of the average material located between the TPC and TOF and $\lambda^{\rm d}_{\rm I}$ is the hadronic interaction length of (anti)deuterons crossing this material. 
The uncertainty on this efficiency due to the uncertainty on the material thickness is given by

\begin{equation}
\Delta \epsilon = | - \dfrac{1}{\lambda^{\rm d}_{\rm I}}| \times \Delta x \times e^{-\Delta x/\lambda^{\rm d}_{\rm I}} \rightarrow \dfrac{\Delta \epsilon}{\epsilon}=\dfrac{\Delta x}{\lambda^{\rm d}_{\rm I}} .
\label{eq:TRDuncertainty}
\end{equation}

Since (anti)deuterons cannot be identified with high purity using the TPC only for \pt > 1.2 $\mathrm{GeV}/\textit{c}$, the uncertainty on the material thickness $\Delta x$ is calculated using the ratio of the TPC--TOF matching efficiencies of (anti)protons measured in data to that in MC simulations. In data, a clean sample of (anti)protons from (anti)$\Lambda$ decays is obtained by applying topological and invariant-mass selections on the reconstructed tracks of the (anti)$\Lambda$ candidate. 

The ratio of the matching efficiencies of (anti)protons measured in data to that in MC simulations is given by 
\begin{equation}
r = \frac{\mathrm{exp} (-x_{\rm true}/\lambda^{\rm p}_{\rm I})}{\mathrm{exp} (-x_{\rm MC}/\lambda^{\rm p}_{\rm I})}  = \mathrm{exp} \left( -\frac{x_{\rm true}-x_{\rm MC}}{\lambda^{\rm p}_{\rm I}}\right) = \mathrm{exp} \left( -\frac{\Delta x}{\lambda^{\rm p}_{\rm I}}\right) ,
\end{equation}

where $x_{\rm true}$ is the "true" material thickness, $x_{\rm MC}$ is its value implemented in the simulation and $\lambda^{\rm p}_{\rm I}$ is the hadronic interaction length of (anti)protons. The latter depends on the inelastic cross section which is very well reproduced by GEANT4 for (anti)protons~\cite{antideuteronInelCS}. 

Finally, the relative systematic uncertainty on the (anti)deuteron \pt spectrum due to the uncertainty on the material thickness between the TPC and TOF (Eq.~\ref{eq:TRDuncertainty}) is calculated as

\begin{equation}
\dfrac{\Delta x}{\lambda^{\rm d}_{\rm I}}=\dfrac{\Delta x}{\lambda^{\rm p}_{\rm I}}\times \dfrac{\sigma_{\rm d}}{\sigma_{\rm p}} ,
\end{equation}

where $\sigma_{\rm d}$ and $\sigma_{\rm p}$ are the inelastic cross sections taken from GEANT4. This uncertainty is found to be about 4.2$\%$ (5.2$\%$) at high \pt for deuterons (antideuterons). 

In the (anti)proton spectrum for \pt > 0.8 $\mathrm{GeV}/\textit{c}$, the uncertainty related to the TPC--TOF matching efficiency is estimated as the difference between data and MC for the TOF matching efficiency for inclusive particles, as a function of \pt, resulting in an average systematic uncertainty of 1$\%$.

The difference between the ratio $\epsilon_{\rm signal}/\epsilon_{\rm event}$ and unity is found to be 1$\%$ at most for (anti)nuclei and 3$\%$ for (anti)protons, and is taken as systematic uncertainty for the signal and event loss correction, as done in Ref.~\cite{Acharya:2020sfy}.

\section{Results} 
\label{sec:Results}

\subsection{Transverse momentum spectra}
The corrected transverse momentum spectra of nuclei and antinuclei are found to be consistent within the uncertainties in all multiplicity classes, as expected in the case of vanishing baryochemical potential at midrapidity. 

\begin{figure}[!hbt]
\centering
\includegraphics[width=0.32\textwidth]{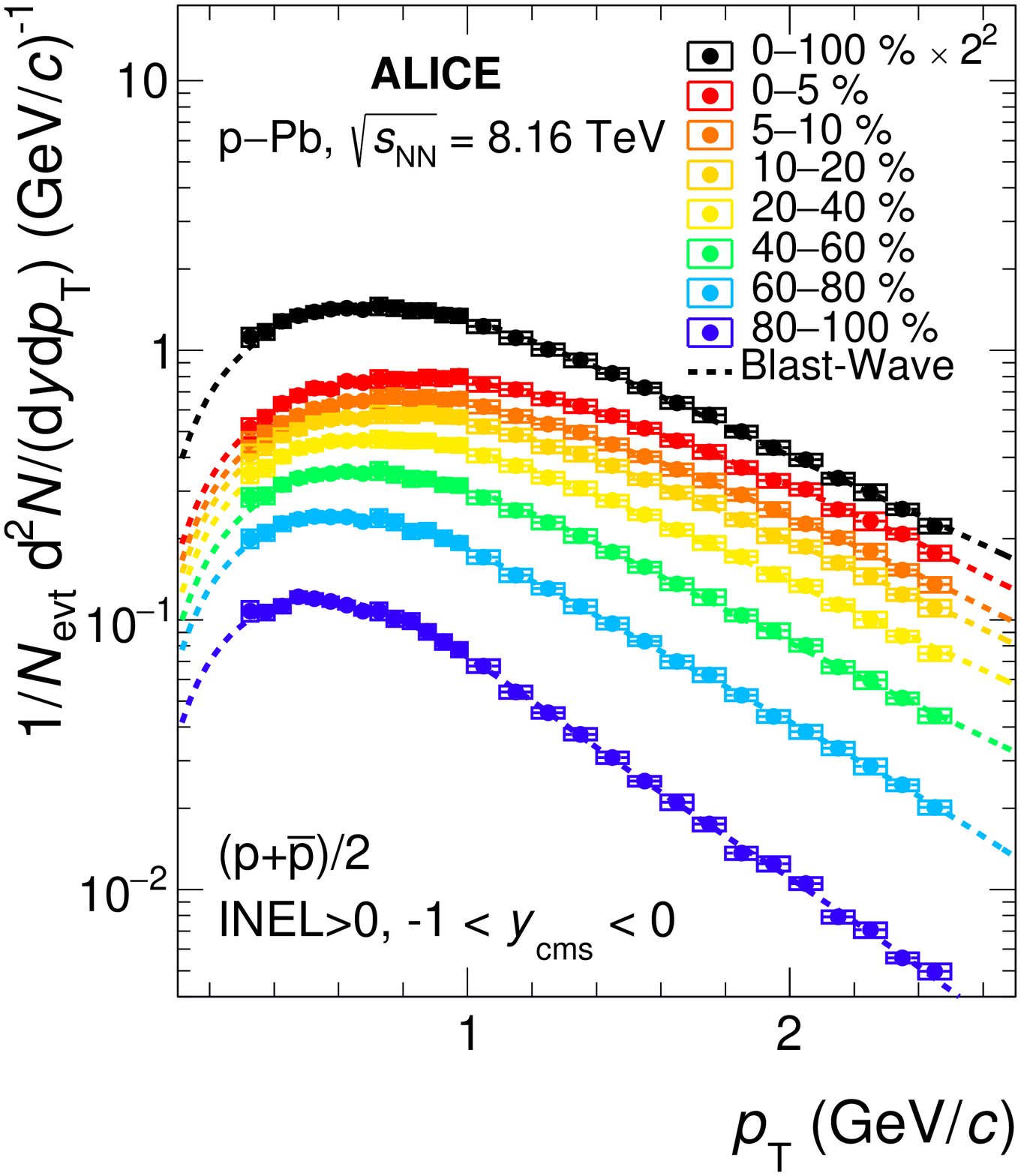}
\includegraphics[width=0.32\textwidth]{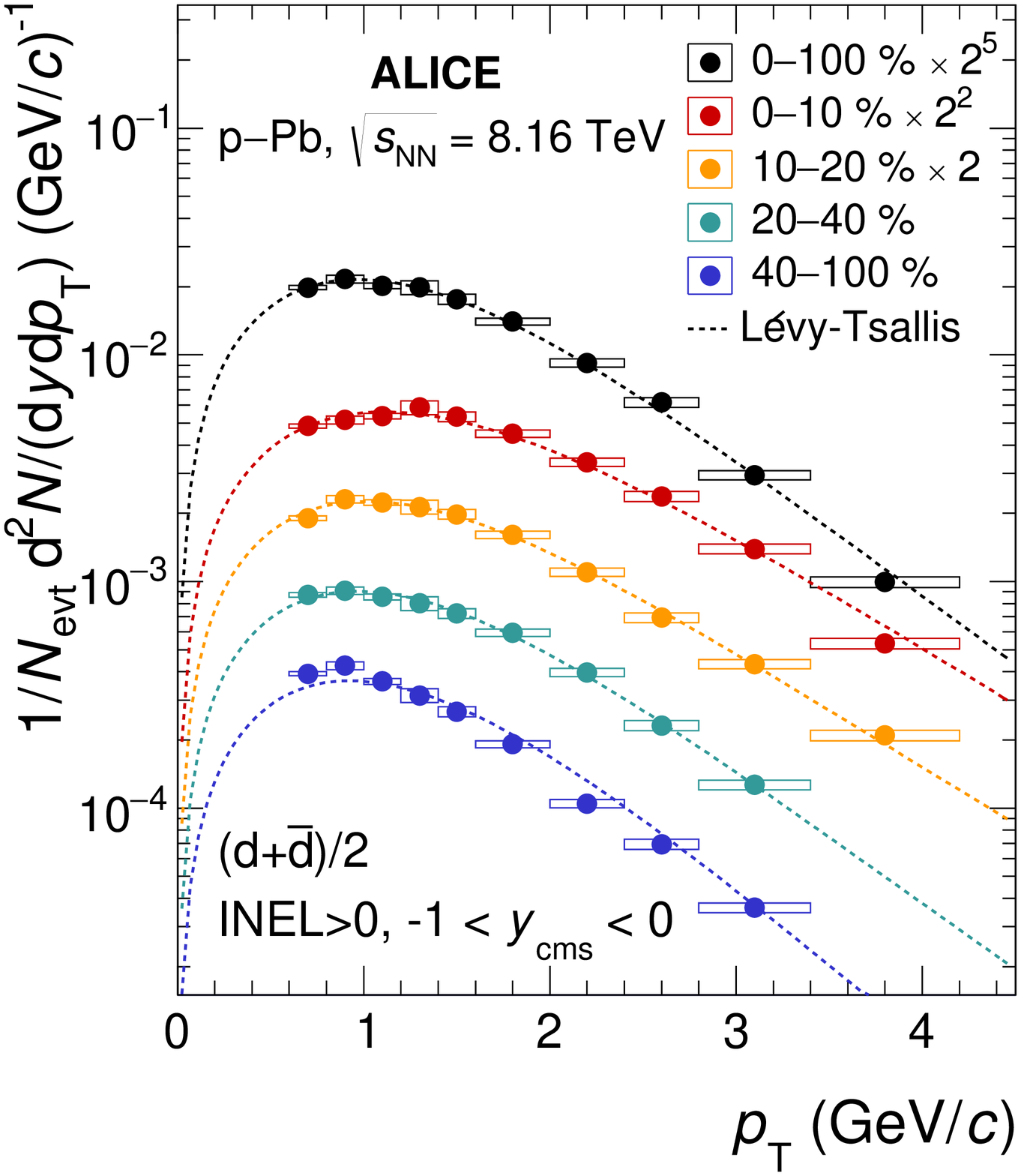}
\includegraphics[width=0.32\textwidth]{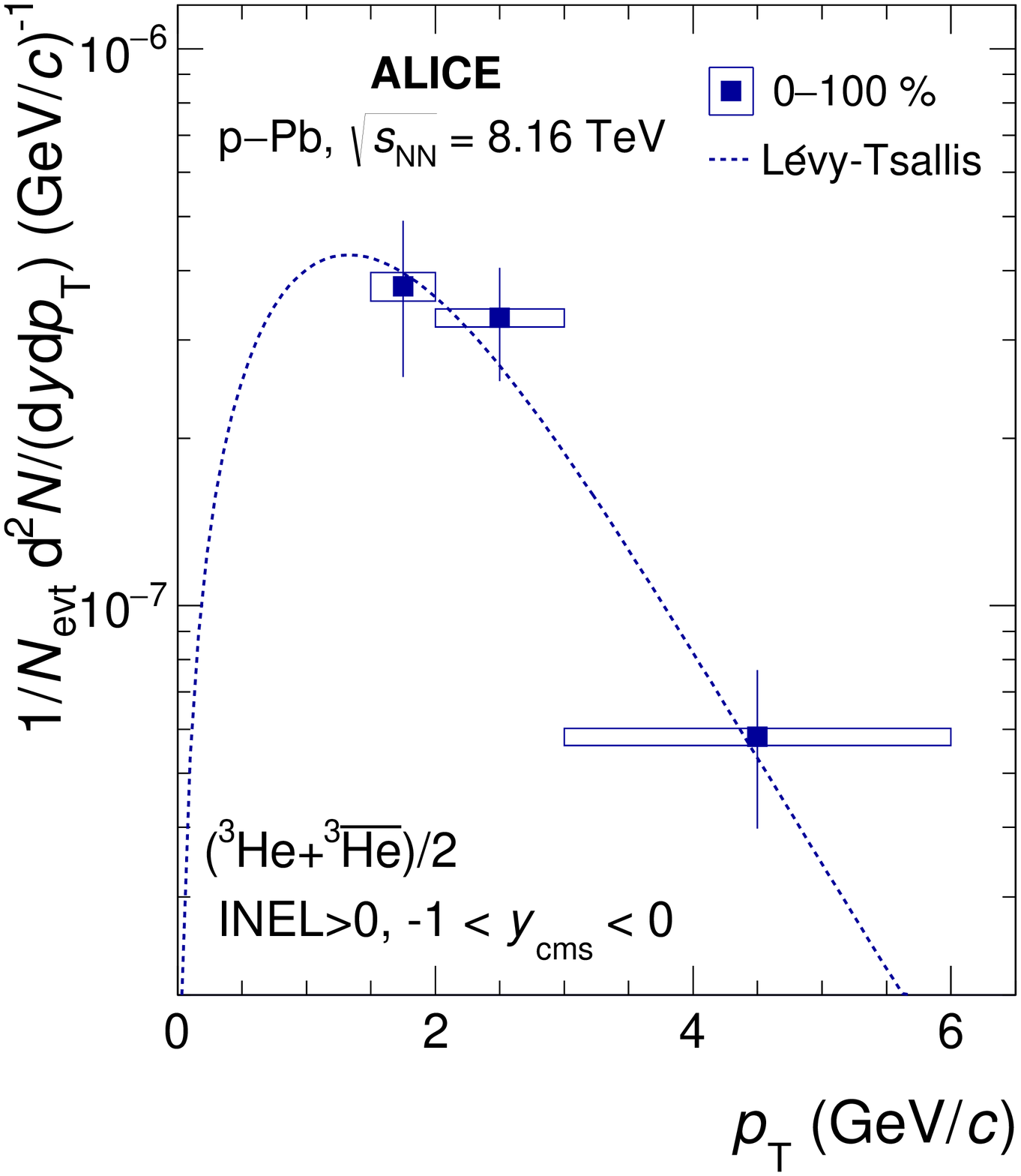}
\caption{Transverse momentum spectra of the average of protons and antiprotons (on the left) and of the average of deuterons and antideuterons (in the middle), in different multiplicity classes, and average of $^3\mathrm{He}$ and $^3\overline{\mathrm{He}}$ (on the right). The \pt distributions of deuterons and $^3$He are fitted using the Lévy--Tsallis function~\cite{Tsallis:1987eu}, while the \pt distributions of protons are fitted using the Blast-Wave fuction~\cite{BlastWave1}. Vertical bars and boxes represent statistical and systematic uncertainties, respectively.}
\label{fig:deuteron_pTspectra}
\end{figure}

Since the ratio between matter and antimatter is compatible with unity, the average \pt spectra of protons (deuterons) and antiprotons (antideuterons) are obtained, as presented in the left (central) panel of Fig.~\ref{fig:deuteron_pTspectra} for several multiplicity classes. Due to the limited statistics of nuclei with mass number $A$ > 2, the transverse momentum spectrum of (anti)$^3$He is measured in the integrated multiplicity class (0--100$\%$) and shown in the right panel of Fig.~\ref{fig:deuteron_pTspectra}. The proton and deuteron \pt distributions become harder with increasing multiplicity. Such behavior was already observed in p--Pb collisions at \snn = 5.02 TeV~\cite{Acharya:2019rys} and in Pb--Pb collisions at \snn = 2.76 TeV~\cite{Acharya:2017dmc}. However, the hardening of the spectra in p--Pb at \snn = 8.16 TeV and that in p--Pb collisions at \snn = 5.02 TeV are different. In the former case, the hardening is more pronounced, for equal centrality classes. The stronger hardening of the spectra is reflected in a larger mean transverse momentum for larger collision energies, which increases of about 35\% from p--Pb collisions at \snn = 5.02 to \snn = 8.16 TeV.
The observed increase of the mean transverse momentum with increasing multiplicity and collision energy could be interpreted either in terms of a collective expansion of the system created in p--Pb collisions or attributed to an increasing contribution of (anti)nuclei production in jets~\cite{Acharya:2019mzb, Acharya:2020ogl, ALICE:2022jmr}, as proposed in Ref.~\cite{Acharya:2019xmu}. The spectra of (anti)nuclei in jets are, indeed, harder than the corresponding spectra in the underlying event, as shown in Refs.~\cite{Acharya:2020ogl, ALICE:2022jmr}, and correspondingly have larger mean transverse momenta.

For the calculation of the \pt-integrated yields (d$N$/d$y$), the Lévy-Tsallis functional form~\cite{Tsallis:1987eu} is used as the default function to fit the (anti)deuteron and (anti)$^{3}$He spectra, while the Blast-Wave~\cite{BlastWave1} function is used as default for (anti)protons, in order to extrapolate to the unmeasured \pt regions (see Table~\ref{tab:YieldDeuteron}).

In the propagation of the systematic uncertainties associated with the \pt spectra to the integrated yields, \pt-correlated and uncorrelated uncertainties have been treated differently. Thanks to a better knowledge of the detectors, a reduction of the uncertainties by a factor of about 3 with respect to previous similar results~\cite{Acharya:2019rys} was possible.
The uncertainties on the relative contribution of secondary nuclei, material budget, signal extraction, hadronic interaction, ITS--TPC matching efficiency, and TPC--TOF matching efficiency for (anti)d, are found to be highly correlated in \pt and are considered as fully \pt correlated, whereas the uncertainties due to track selection and signal-loss efficiency, and to $^{3}$H contamination subtraction for (anti)$^3$He, are found to be mostly uncorrelated with transverse momentum. 
Data points in the spectra are shifted up and down by the correlated part of the systematic uncertainties and refitted to provide extrapolated values. Half of the difference between the two resulting values has been used as the \pt correlated part of the systematic uncertainty.
To evaluate the \pt uncorrelated contribution to the total systematic uncertainty of the yield, the Gaussian sampling method is applied. The latter consists in shifting the average (anti)nucleus data points using a random Gaussian distribution centered at the measured value of each \pt interval, with a standard deviation given by the \pt uncorrelated systematic uncertainty of each point. 
The obtained \pt spectra are fitted with several functions, thus yielding various d$N$/d$y$ values. The RMS of the distribution of such integrated yield values is assigned as systematic uncertainty.
The contribution given by the spread between the values obtained by different fit functions is also considered to estimate the systematic uncertainty on the yield. The spectra are refitted using several functions, namely Boltzmann~\cite{BHALLA1981446}, Fermi-Dirac~\cite{fermidirac}, $m_{\mathrm T}$-exponential~\cite{Adler:2003cb}, Blast-Wave~\cite{BlastWave1}, and Lévy-Tsallis~\cite{Tsallis:1987eu} when not used as default, and, for (anti)deuterons only, the power law~\cite{Albajar:1989an} functional form. For this contribution, half of the difference between the maximum and the minimum yield is taken as systematic uncertainty. All the discussed contributions were finally summed in quadrature. 

Based on the fit functions, also the mean transverse momentum $\langle p_{\rm T}\rangle$ of average of (anti)$^3$He, and average of (anti)deuteron, and average of (anti)proton is calculated, in the latter cases for several multiplicity classes. The results are reported in Table~\ref{tab:MeanpT}.

\begin{table}[!hbt]
\centering
\caption{Integrated yields (d$N$/d$y$) of (anti)protons, (anti)deuterons, and (anti)$^{3}$He for each multiplicity class. The first uncertainty is statistical and the second is the systematic one. }
\begin{tabular}{ccccc}
\toprule
V0A class & $\langle$d$N_{\rm ch}$/d$\eta_{\mathrm {lab}}$$\rangle \big|_{|\eta_{\mathrm {lab}}|<0.5}$ & d$N$/d$y$ [(p+$\overline{\mathrm p}$)/2] & d$N$/d$y$ [(d+$\overline{\mathrm d}$)/2] & d$N$/d$y$ [($^{3}$He+$^{3}\overline{\mathrm{He}}$)/2]\\
 & & $\times  10^{-1}$ & $\times  10^{-3}$ &$\times  10^{-6}$ \\
\midrule
0--100$\%$    & 20.3 $\pm$ 0.6 & 5.51 $\pm$ 0.02 $\pm$ 0.15 & 1.27 $\pm$ 0.01 $\pm$ 0.04 & 1.15 $\pm$ 0.16  $\pm$ 0.13\\
0--10$\%$     & 47.8 $\pm$ 1.2  & & 3.11 $\pm$ 0.03 $\pm$ 0.10 & \\
0--5$\%$ & 53.2 $\pm$ 1.4 & 13.51 $\pm$ 0.04 $\pm$ 0.36 & &\\
5--10$\%$ & 42.4 $\pm$ 1.1 & 10.99 $\pm$ 0.04 $\pm$ 0.29 & &\\
10--20$\%$   & 35.5 $\pm$ 0.9 & 9.37 $\pm$ 0.03 $\pm$ 0.25 & 2.31 $\pm$ 0.02 $\pm$ 0.07 & \\
20--40$\%$   & 26.9 $\pm$ 0.7 & 7.27 $\pm$ 0.02 $\pm$ 0.19 & 1.72 $\pm$ 0.02 $\pm$ 0.05 & \\
40--100$\%$ & 13.0 $\pm$ 0.4  & & 0.64 $\pm$ 0.01 $\pm$ 0.02 & \\
40--60$\%$ & 18.4 $\pm$ 0.5 & 5.06 $\pm$ 0.02 $\pm$ 0.13 & &\\
60--80$\%$ & 11.0 $\pm$ 0.3 & 3.06 $\pm$ 0.01 $\pm$ 0.08 & &\\
80--100$\%$ & 4.5 $\pm$ 0.1 & 1.27 $\pm$ 0.01 $\pm$ 0.03 & &\\
\bottomrule
\end{tabular}
\label{tab:YieldDeuteron}
\end{table}

\begin{table}
\centering
\caption{Mean transverse momentum of the average (anti)proton and (anti)deuteron spectra for each multiplicity class and of the average (anti)$^3$He spectrum. The first uncertainty is statistical and the second is the systematic one. The mean \pt is obtained from the Lévy--Tsallis fit for the average (anti)deuteron and (anti)$^3$He spectra, whereas the Blast-Wave fit is used for the (anti)proton case. }
\begin{tabular}{cccc}
\toprule
V0A class & proton $\langle p_{\rm T}\rangle$ ($\mathrm{GeV}/\textit{c}$) & deuteron $\langle p_{\rm T}\rangle$ ($\mathrm{GeV}/\textit{c}$) & $^{3}$He $\langle p_{\rm T}\rangle$ ($\mathrm{GeV}/\textit{c}$)\\
\midrule
   0--100$\%$	& 1.176 $\pm$ 0.006 $\pm$ 0.035 &  1.45 $\pm$ 0.02 $\pm$ 0.02   &   2.08 $\pm$ 0.20 $\pm$ 0.06\\
   0--10$\%$	&	&  1.70 $\pm$ 0.02 $\pm$ 0.03\\
   0--5$\%$	    & 1.286 $\pm$ 0.006 $\pm$ 0.039  & & \\
   5--10$\%$	& 1.252 $\pm$ 0.006 $\pm$ 0.038  & & \\
  10--20$\%$ 	& 1.236 $\pm$ 0.006 $\pm$ 0.037 &  1.59 $\pm$ 0.02 $\pm$ 0.03\\
  20--40$\%$ 	&  1.194 $\pm$ 0.006 $\pm$ 0.036  &  1.46  	$\pm$ 0.02 $\pm$ 0.02\\
 40--100$\%$	&    &  1.36  	$\pm$ 0.02 $\pm$ 0.02\\
    40--60$\%$	    & 1.115 $\pm$ 0.005 $\pm$ 0.033 & & \\
    60--80$\%$	    & 0.983 $\pm$ 0.005 $\pm$ 0.029 & & \\
    80--100$\%$	    & 0.839 $\pm$ 0.004 $\pm$ 0.025 & & \\
\bottomrule
\end{tabular}
\label{tab:MeanpT}
\end{table}

\subsection{Coalescence parameters}
According to coalescence models, the production of light (anti)nuclei can be explained via the coalescence of protons and neutrons which are close in phase space at the freeze-out and match the spin, thus forming a nucleus~\cite{Coalescence3}. The key parameter of the coalescence models is the coalescence probability, $B_A$, which is experimentally accessible using the invariant yields of protons and that of nuclei, following Eq.~\ref{eq:BA}.

\begin{figure}[!hbt]
\centering
\begin{minipage}[b]{0.49\textwidth}
    \includegraphics[width=\textwidth]{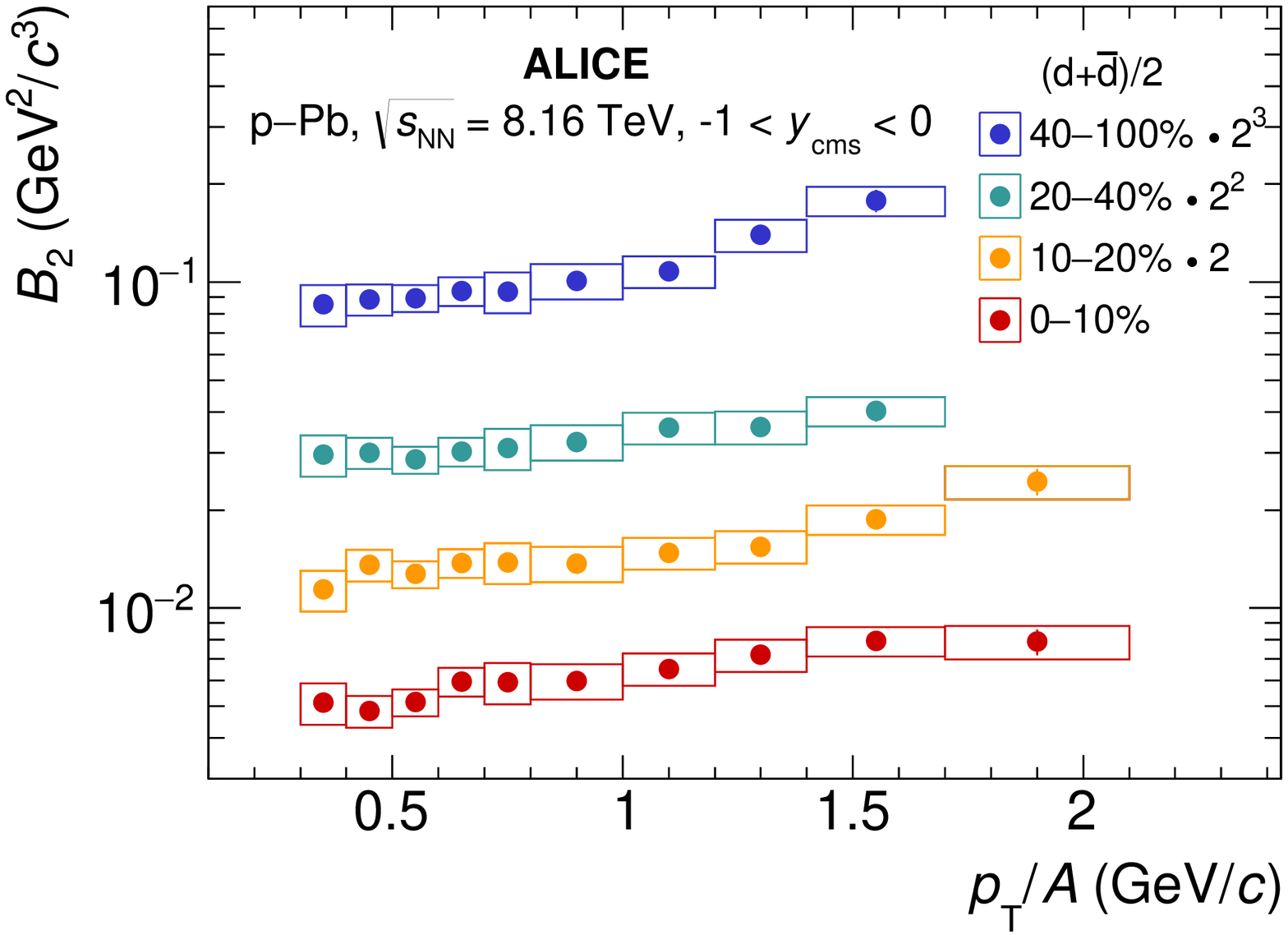}
\end{minipage}   
\hfill
\begin{minipage}[b]{0.49\textwidth}
	\includegraphics[width=\textwidth]{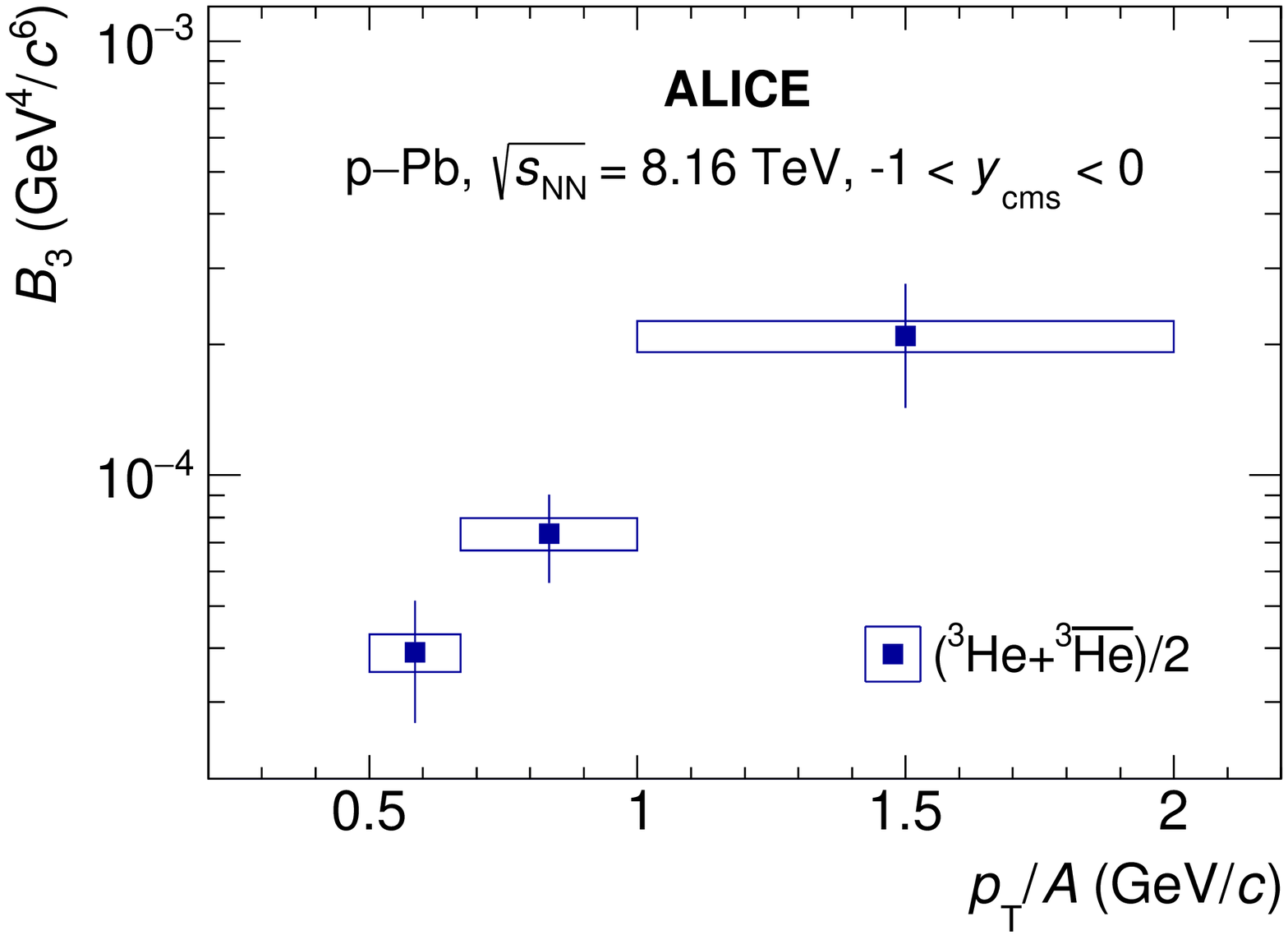}
\end{minipage}
\caption{Coalescence parameters $B_2$ (left panel) and $B_3$ (right panel) as a function of $p_{\mathrm T}$/$A$, measured for deuterons and $^3$He, respectively. Statistical uncertainties are represented as vertical lines whereas boxes represent the systematic ones. }
\label{fig:CoalescenceParameters}
\end{figure}

In Fig.~\ref{fig:CoalescenceParameters} the coalescence parameters $B_2$ and $B_3$ are shown as a function of the transverse momentum per nucleon (\pt/$A$) for different multiplicity classes. The \pt-differential yields of protons are averaged in the multiplicity classes and in the \pt intervals of the deuteron and $^3$He analyses, to obtain the coalescence parameters. 
The coalescence parameters $B_2$ and $B_3$ show a rising trend with increasing \pt/$A$ in all multiplicity classes. 
The simple coalescence model, where only momentum correlations are considered, predicts a constant trend of the coalescence parameters with \pt/$A$ for a fixed multiplicity. 
It was already demonstrated that, under the hypothesis of the simple coalescence model, the coalescence parameter develops an increasing trend within a wide multiplicity interval because of the different hardening of the proton and nucleus spectra with multiplicity~\cite{deuteron_pp7TeV}. 
This could explain the increasing trend of $B_2$ in the multiplicity intervals used for this measurement. However, a re-calculation of the coalescence parameter $B_3$ measured in p--Pb collisions at \snn = 5.02 TeV under the hypothesis of the simple coalescence model did not reproduce the experimental data~\cite{Acharya:2019xmu}. This implies that the increase of $B_3$ with \pt/$A$ cannot be fully explained by this kinematic effect and it is therefore a genuine physical effect. As a further confirmation, a rising trend is also observed in very narrow multiplicity intervals for both $B_2$ and $B_3$ in high-multiplicity pp collisions at $\sqrt{s}$ = 13 TeV~\cite{nuclei_pp_13TeV}.

\begin{figure}[!hbt]
\centering
	\includegraphics[width=0.7\textwidth]{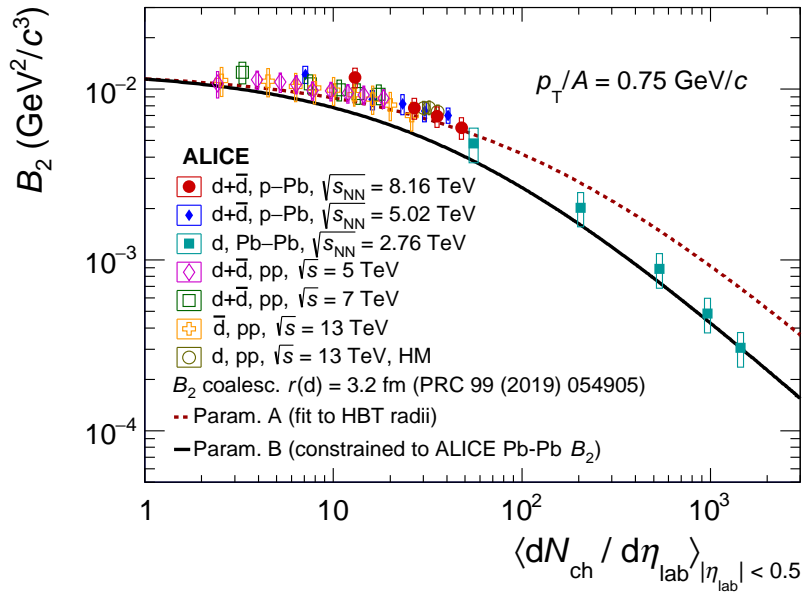}
\caption{$B_2$ as a function of the mean charged-particle multiplicity density for a fixed value of \mbox{$p_{\mathrm T}$/$A$ = 0.75 GeV/$c$}. The experimental results are compared to the coalescence calculations from Ref.~\cite{CoalescenceTheory} using two different parameterizations for the system size as a function of the mean charged-particle multiplicity density. }
\label{fig:B2_vs_mult}
\end{figure}

In order to investigate the dependence of the coalescence probability on the size of the particle emitting source, as suggested by state-of-the-art coalescence models~\cite{BRAUNMUNZINGER2019144,CoalescenceTheory,Mrowczynski:2020ugu}, the $B_2$ parameters extracted in several collision systems and LHC energies~\cite{deuteron_pp7TeV, Acharya:2019rys, Adam:2015vda, Acharya:2017fvb, nuclei_pp_13TeV, nuclei_pp_5TeV, Acharya:2020sfy, Acharya:2019xmu, helium3_flow_5TeV, ALICE:2013wgn, ALICE:2013mez, ALICE:2020nkc, ALICE:2018pal} are compared as a function of the charged-particle multiplicity for a fixed value of $p_{\mathrm T}$/$A$ in Fig.~\ref{fig:B2_vs_mult}.
The measurements show a smooth transition from low to high charged-particle multiplicity densities, which correspond to an increasing system size. 
The continuous and consistent trend of $B_{2}$ with increasing multiplicity suggests that the dominant production mechanism evolves smoothly as a function of the system size and is independent of the collision system and center-of-mass energy. The experimental results are compared to the theoretical calculations from coalescence~\cite{CoalescenceTheory}, using two different parameterizations of the size of the source as a function of multiplicity. 
Parameterization A is based on a fit of the source radii measured by ALICE as a function of multiplicity using femtoscopic techniques~\cite{Abelev:2014pja}. 
Parameterization B, instead, uses parameters constrained to reproduce the $B_2$ of deuterons as extracted by ALICE in central (0--10$\%$) Pb--Pb collisions at \snn = 2.76 TeV. 
Parameterization B is less favoured by the p--Pb at \snn = 8.16 TeV results with respect to the parameterization A.
Future dedicated studies of the relation between source size and multiplicity (and as a function of \pt) will be crucial to further constrain or test the coalescence models.

\subsection{Ratio to protons}

\begin{figure}[!hbt]
\centering
\begin{minipage}[b]{0.49\textwidth}
	\includegraphics[width=\textwidth]{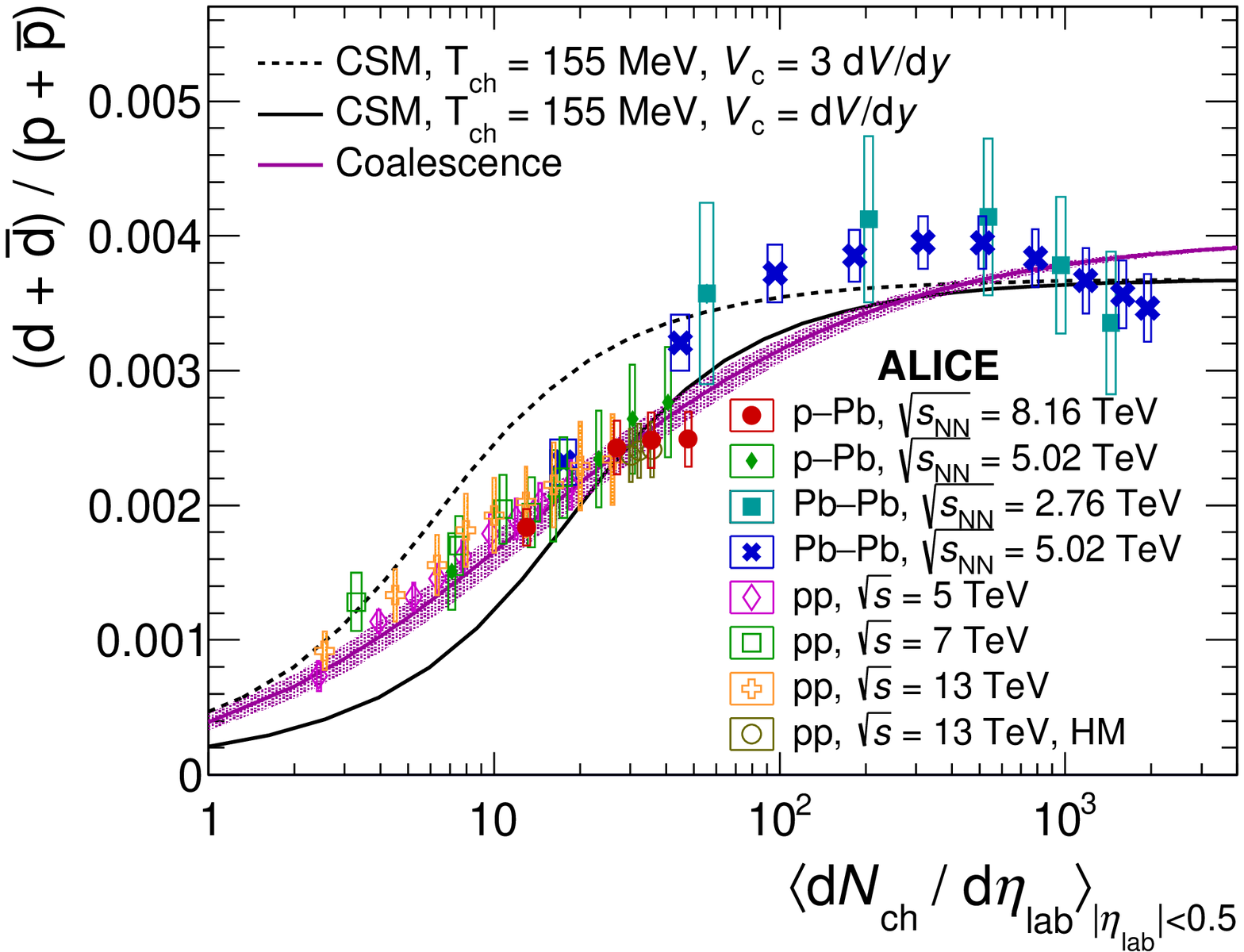}
\end{minipage}
  \hfill
\begin{minipage}[b]{0.49\textwidth}
    \includegraphics[width=\textwidth]{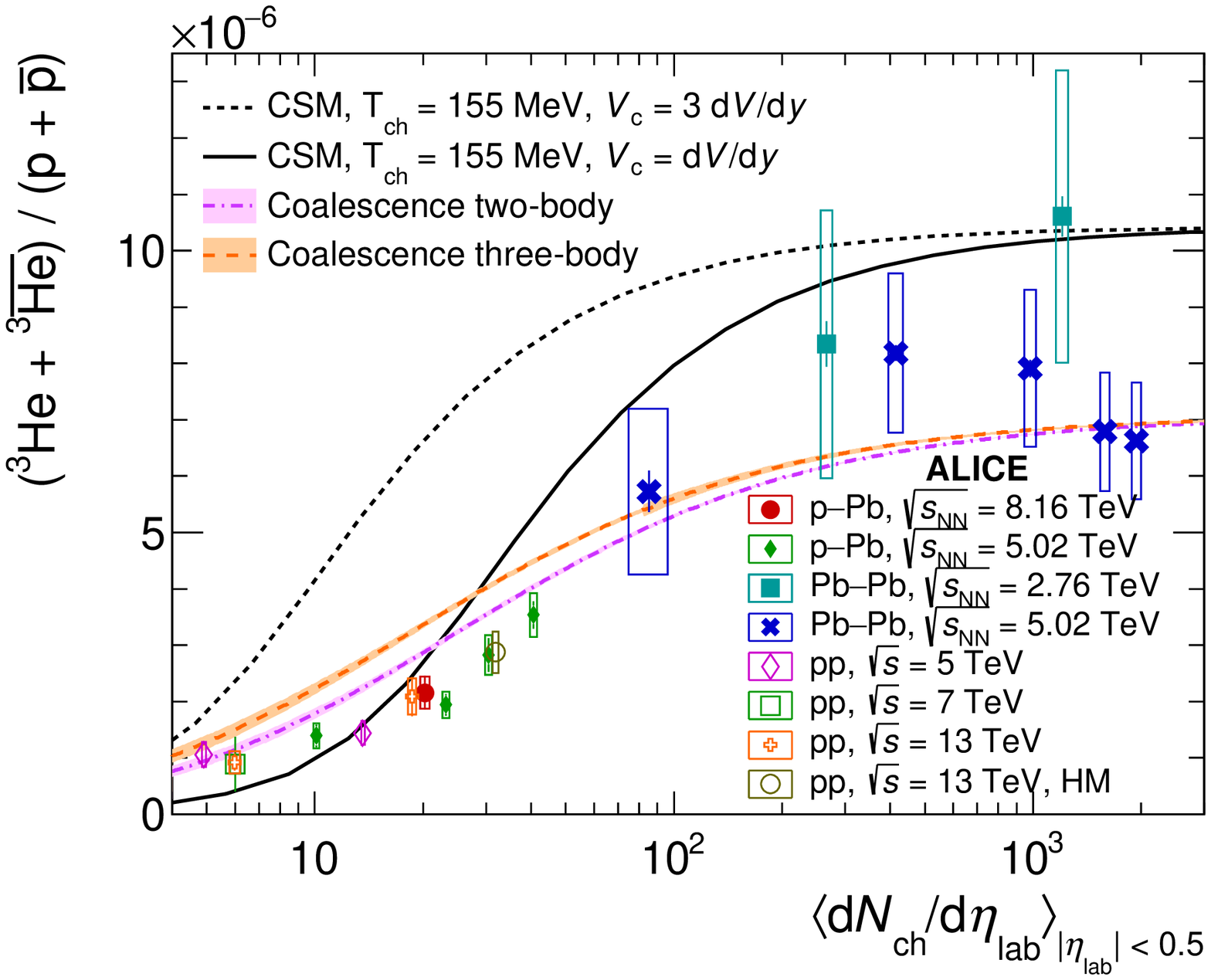}
\end{minipage}   
\caption{Ratio of deuteron (left panel) and of (anti)$^3$He (right panel) and proton production yields as a function of the charged-particle multiplicity in different collision systems and energies. Statistical uncertainties are represented as vertical lines, whereas boxes represent the systematic ones. The results are compared with the expectations of CSM and Coalescence models. }
\label{fig:NucleiOverP}
\end{figure}

The ratio of the measured yields of nuclei and that of protons is also sensitive to the light nuclei production mechanism. In Fig.~\ref{fig:NucleiOverP} the yield ratios to protons for deuterons (left panel), and $^3$H or $^3$He (right panel) measured in pp, p--Pb and Pb--Pb collisions as a function of $\langle$d$N_{\mathrm{ch}}/$d$\eta_{\mathrm{lab}}\rangle$~\cite{deuteron_pp7TeV, Acharya:2019rys, Adam:2015vda, Acharya:2017fvb, nuclei_pp_13TeV, nuclei_pp_5TeV, Acharya:2020sfy, Acharya:2019xmu, helium3_flow_5TeV, ALICE:2013wgn, ALICE:2013mez, ALICE:2020nkc, ALICE:2018pal,pbpb5TeV} are compared with the expectations of the models. In the canonical statistical hadronization model (CSM) used here, exact conservation of baryon number (B), electric charge (Q) and strangeness (S) is implemented in the so-called correlation volume $V_{\mathrm{c}}$. Different values of $V_{\mathrm{c}}$ are tested, extending from one to three units of rapidity. 
Considering that the matter produced at midrapidity at the LHC is practically baryon free, the model is applied for exactly vanishing values of the conserved charges \mbox{B = Q = S = 0}. The system is assumed to be in full chemical equilibrium and the chemical freezeout temperature is fixed at 155 MeV, independent of multiplicity~\cite{Vovchenko:2018fiy}. Recent developments of the CSM, called $\gamma_{\mathrm{S}}$CSM, include an incomplete equilibration of strangeness, described by the strangeness saturation parameter $\gamma_{\mathrm{S}}$, a multiplicity-dependent chemical freezeout temperature and a correlation volume extending over three units of rapidity~\cite{vanillaCSM}. The $\gamma_{\mathrm{S}}$CSM model reproduces quite well the measured hadron-to-pion ratios as a function of multiplicity, except for the p/$\pi$ ratio which is systematically overestimated by the model approximately by 2$\sigma$. No prediction is currently available for the nuclei-to-proton ratio from the $\gamma_{\mathrm{S}}$CSM model.

In the coalescence calculations, the (anti)nuclei formation probability is given by the overlap of the nucleon phase-space distributions in the emission source with the Wigner density of the bound state. The latter is calculated using a Gaussian approximation for the (anti)nuclei internal wave function~\cite{coalescenceSmallSystems}. In the case of $A=3$ nuclei, predictions from both two-body and three-body coalescence are considered~\cite{coalescenceSmallSystems}. In the two-body coalescence of $^{3}$He ($^{3}$H), a two-step process is assumed: first, the deuteron is formed by the coalescence of a proton and a neutron, then the $^{3}$He ($^{3}$H) is formed by the coalescence of the deuteron and a proton (neutron). In the three-body coalescence, three nucleons form $^{3}$He ($^{3}$H) at once. 

A smooth increase of deuteron-to-proton yield ratio (d/p) and  $^3$He-to-proton yield ratio ($^3$He/p) with the system size is observed, reaching constant values in Pb--Pb collisions. The plateau at high multiplicities is described by the grand-canonical statistical model~\cite{SHM1, CoalescenceTheory, Coalescence2}. 
The two ratios show a similar trend with $\langle$d$N_{\mathrm{ch}}/$d$\eta_{\mathrm{lab}}\rangle$, however the increase from pp to Pb--Pb results is about a factor of 3 larger for $^3$He/p than for d/p. 
The evolution of the d/p ratio is well described by the coalescence approach over the full multiplicity interval covered by the existing measurements. 
The CSM calculations asymptotically converge towards the grand-canonical limit at high multiplicity and they are both consistent with the Pb--Pb measurements at \snn = 2.76 TeV within the uncertainties. At low and intermediate multiplicities, these calculations provide only a qualitative description of this ratio, with the version using as correlation volume $V_{\rm c}$ = d$V$/d$y$ being consistent with the data at intermediate multiplicity and the version using $V_{\rm c}$ = 3d$V$/d$y$ at low multiplicity only.

The $^3$He/p ratio, shown in Fig.~\ref{fig:NucleiOverP}, is fairly well described by the coalescence approach, especially at low and high charged-particle multiplicity densities, where the compatibility between data and two-body coalescence calculations is within 1.5$\sigma$. Some tension is observed at intermediate multiplicities \mbox{(10 $<\langle$d$N_{\mathrm{ch}}/$d$\eta_{\mathrm{lab}}\rangle<$ 40)}, where the predictions underestimate the data, up to 5$\sigma$. 
Similarly to the d/p ratio, the CSM calculations provide a qualitative description of the data, with a compatibility that, in the case of the CSM with $V_{\rm c}$ = d$V$/d$y$, decreases with increasing multiplicities, from 2$\sigma$ to 5$\sigma$. The version of CSM with $V_{\rm c}$ = 3d$V$/d$y$ is excluded by up to 12$\sigma$ at low and intermediate multiplicities, while in the grand-canonical limit the agreement reaches 4$\sigma$. 

\section{Summary} 
\label{sec:Conclusions}

Measurements of (anti)proton, (anti)deuteron, and (anti)$^{3}$He production in p--Pb collisions at \snn = 8.16 TeV are presented. These results contribute to the understanding of the light (anti)nuclei production mechanism complementing the existing picture, which includes measurements done in different collision systems and at different center-of-mass energies.  

A hardening of the (anti)proton and (anti)deuteron \pt spectra with increasing event multiplicity and collision energy is observed, consistently with previous measurements~\cite{Acharya:2019rys, Acharya:2019xmu, Acharya:2020sfy, deuteron_pp7TeV}, and could be interpreted either in terms of a collective expansion of the system created in p--Pb collisions or attributed to an increasing contribution of (anti)nuclei production in jets~\cite{Acharya:2019mzb, Acharya:2020ogl, ALICE:2022jmr}, as proposed in Ref.~\cite{Acharya:2019xmu}. 
The production mechanisms of (anti)deuterons and (anti)$^{3}$He are investigated by comparing the multiplicity dependence of the coalescence parameters $B_{A}$ and their yields relative to protons, with the predictions of the canonical statistical model and of the coalescence model. A smooth evolution of these observables with multiplicity across different collision systems and energies is seen. The intermediate multiplicity range, which is covered in this measurement, is particularly interesting as it links existing results in pp and Pb--Pb collisions, corresponding to small and large system sizes, respectively. The results of $B_2$ as a function of the mean charged-particle multiplicity density show a good agreement with the coalescence model that uses the parameterization of the source radii based on femtoscopic techniques. 

The results presented in this article corroborate a better description of the deuteron production measurements by the coalescence model than by the canonical statistical hadronization model.
However, in the case of the $^{3}$He/p ratio, significant deviations are measured at intermediate multiplicities between the data and the predictions of both the two-body and the three-body coalescence calculations.
Implementations of the CSM with a fixed correlation volume cannot describe quantitatively the nucleus-to-proton ratio in the full multiplicity range spanned by the ALICE measurements, but capture the increasing trend qualitatively. At high multiplicity, CSM calculations are consistent with the data in the grand canonical limit.

\newenvironment{acknowledgement}{\relax}{\relax}
\begin{acknowledgement}
\section*{Acknowledgements}

The ALICE Collaboration would like to thank all its engineers and technicians for their invaluable contributions to the construction of the experiment and the CERN accelerator teams for the outstanding performance of the LHC complex.
The ALICE Collaboration gratefully acknowledges the resources and support provided by all Grid centres and the Worldwide LHC Computing Grid (WLCG) collaboration.
The ALICE Collaboration acknowledges the following funding agencies for their support in building and running the ALICE detector:
A. I. Alikhanyan National Science Laboratory (Yerevan Physics Institute) Foundation (ANSL), State Committee of Science and World Federation of Scientists (WFS), Armenia;
Austrian Academy of Sciences, Austrian Science Fund (FWF): [M 2467-N36] and Nationalstiftung f\"{u}r Forschung, Technologie und Entwicklung, Austria;
Ministry of Communications and High Technologies, National Nuclear Research Center, Azerbaijan;
Conselho Nacional de Desenvolvimento Cient\'{\i}fico e Tecnol\'{o}gico (CNPq), Financiadora de Estudos e Projetos (Finep), Funda\c{c}\~{a}o de Amparo \`{a} Pesquisa do Estado de S\~{a}o Paulo (FAPESP) and Universidade Federal do Rio Grande do Sul (UFRGS), Brazil;
Ministry of Education of China (MOEC) , Ministry of Science \& Technology of China (MSTC) and National Natural Science Foundation of China (NSFC), China;
Ministry of Science and Education and Croatian Science Foundation, Croatia;
Centro de Aplicaciones Tecnol\'{o}gicas y Desarrollo Nuclear (CEADEN), Cubaenerg\'{\i}a, Cuba;
Ministry of Education, Youth and Sports of the Czech Republic, Czech Republic;
The Danish Council for Independent Research | Natural Sciences, the VILLUM FONDEN and Danish National Research Foundation (DNRF), Denmark;
Helsinki Institute of Physics (HIP), Finland;
Commissariat \`{a} l'Energie Atomique (CEA) and Institut National de Physique Nucl\'{e}aire et de Physique des Particules (IN2P3) and Centre National de la Recherche Scientifique (CNRS), France;
Bundesministerium f\"{u}r Bildung und Forschung (BMBF) and GSI Helmholtzzentrum f\"{u}r Schwerionenforschung GmbH, Germany;
General Secretariat for Research and Technology, Ministry of Education, Research and Religions, Greece;
National Research, Development and Innovation Office, Hungary;
Department of Atomic Energy Government of India (DAE), Department of Science and Technology, Government of India (DST), University Grants Commission, Government of India (UGC) and Council of Scientific and Industrial Research (CSIR), India;
National Research and Innovation Agency - BRIN, Indonesia;
Istituto Nazionale di Fisica Nucleare (INFN), Italy;
Japanese Ministry of Education, Culture, Sports, Science and Technology (MEXT) and Japan Society for the Promotion of Science (JSPS) KAKENHI, Japan;
Consejo Nacional de Ciencia (CONACYT) y Tecnolog\'{i}a, through Fondo de Cooperaci\'{o}n Internacional en Ciencia y Tecnolog\'{i}a (FONCICYT) and Direcci\'{o}n General de Asuntos del Personal Academico (DGAPA), Mexico;
Nederlandse Organisatie voor Wetenschappelijk Onderzoek (NWO), Netherlands;
The Research Council of Norway, Norway;
Commission on Science and Technology for Sustainable Development in the South (COMSATS), Pakistan;
Pontificia Universidad Cat\'{o}lica del Per\'{u}, Peru;
Ministry of Education and Science, National Science Centre and WUT ID-UB, Poland;
Korea Institute of Science and Technology Information and National Research Foundation of Korea (NRF), Republic of Korea;
Ministry of Education and Scientific Research, Institute of Atomic Physics, Ministry of Research and Innovation and Institute of Atomic Physics and University Politehnica of Bucharest, Romania;
Ministry of Education, Science, Research and Sport of the Slovak Republic, Slovakia;
National Research Foundation of South Africa, South Africa;
Swedish Research Council (VR) and Knut \& Alice Wallenberg Foundation (KAW), Sweden;
European Organization for Nuclear Research, Switzerland;
Suranaree University of Technology (SUT), National Science and Technology Development Agency (NSTDA), Thailand Science Research and Innovation (TSRI) and National Science, Research and Innovation Fund (NSRF), Thailand;
Turkish Energy, Nuclear and Mineral Research Agency (TENMAK), Turkey;
National Academy of  Sciences of Ukraine, Ukraine;
Science and Technology Facilities Council (STFC), United Kingdom;
National Science Foundation of the United States of America (NSF) and United States Department of Energy, Office of Nuclear Physics (DOE NP), United States of America.
In addition, individual groups or members have received support from:
Marie Sk\l{}odowska Curie (grant no. 896850), European Union.

\end{acknowledgement}

\bibliographystyle{utphys}   
\bibliography{bibliography}

\newpage
\appendix

%
%

\section{The ALICE Collaboration}
\label{app:collab}
\begin{flushleft} 
\small

S.~Acharya\,\orcidlink{0000-0002-9213-5329}\,$^{\rm 131}$, 
D.~Adamov\'{a}\,\orcidlink{0000-0002-0504-7428}\,$^{\rm 85}$, 
A.~Adler$^{\rm 69}$, 
J.~Adolfsson\,\orcidlink{0000-0001-5651-4025}\,$^{\rm 74}$, 
G.~Aglieri Rinella\,\orcidlink{0000-0002-9611-3696}\,$^{\rm 32}$, 
M.~Agnello\,\orcidlink{0000-0002-0760-5075}\,$^{\rm 29}$, 
N.~Agrawal\,\orcidlink{0000-0003-0348-9836}\,$^{\rm 49}$, 
Z.~Ahammed\,\orcidlink{0000-0001-5241-7412}\,$^{\rm 131}$, 
S.~Ahmad$^{\rm 15}$, 
S.U.~Ahn\,\orcidlink{0000-0001-8847-489X}\,$^{\rm 70}$
% ...

\section*{Affiliation Notes}

$^{\rm I}$ Deceased\\
$^{\rm II}$ Also at: Italian National Agency for New Technologies, Energy and Sustainable Economic Development (ENEA), Bologna, Italy\\
$^{\rm III}$ Also at: Dipartimento DET del Politecnico di Torino, Turin, Italy\\
$^{\rm IV}$ Also at: Department of Applied Physics, Aligarh Muslim University, Aligarh, India\\
$^{\rm V}$ Also at: Institute of Theoretical Physics, University of Wroclaw, Poland\\
$^{\rm VI}$ Also at: An institution covered by a cooperation agreement with CERN\\
$^{\rm VII}$ Also at: Indian Institute of Science Education and Research (IISER) Berhampur, Odisha, India\\

\section*{Collaboration Institutes}

$^{1}$ A.I. Alikhanyan National Science Laboratory (Yerevan Physics Institute) Foundation, Yerevan, Armenia\\
$^{2}$ AGH University of Krakow, Cracow, Poland\\
$^{3}$ Bogolyubov Institute for Theoretical Physics, National Academy of Sciences of Ukraine, Kiev, Ukraine\\
$^{4}$ Bose Institute, Department of Physics  and Centre for Astroparticle Physics and Space Science (CAPSS), Kolkata, India\\
$^{5}$ California Polytechnic State University, San Luis Obispo, California, United States\\
$^{6}$ Central China Normal University, Wuhan, China\\
$^{7}$ Centro de Aplicaciones Tecnol\'{o}gicas y Desarrollo Nuclear (CEADEN), Havana, Cuba\\
$^{8}$ Centro de Investigaci\'{o}n y de Estudios Avanzados (CINVESTAV), Mexico City and M\'{e}rida, Mexico\\
$^{9}$ Chicago State University, Chicago, Illinois, United States\\
$^{10}$ China Institute of Atomic Energy, Beijing, China\\
$^{11}$ Chungbuk National University, Cheongju, Republic of Korea\\
$^{12}$ Comenius University Bratislava, Faculty of Mathematics, Physics and Informatics, Bratislava, Slovak Republic\\
$^{13}$ COMSATS University Islamabad, Islamabad, Pakistan\\
$^{14}$ Creighton University, Omaha, Nebraska, United States\\
$^{15}$ Department of Physics, Aligarh Muslim University, Aligarh, India\\
$^{16}$ Department of Physics, Pusan National University, Pusan, Republic of Korea\\
$^{17}$ Department of Physics, Sejong University, Seoul, Republic of Korea\\
$^{18}$ Department of Physics, University of California, Berkeley, California, United States\\
$^{19}$ Department of Physics, University of Oslo, Oslo, Norway\\
$^{20}$ Department of Physics and Technology, University of Bergen, Bergen, Norway\\
$^{21}$ Dipartimento di Fisica, Universit\`{a} di Pavia, Pavia, Italy\\
$^{22}$ Dipartimento di Fisica dell'Universit\`{a} and Sezione INFN, Cagliari, Italy\\
$^{23}$ Dipartimento di Fisica dell'Universit\`{a} and Sezione INFN, Trieste, Italy\\
$^{24}$ Dipartimento di Fisica dell'Universit\`{a} and Sezione INFN, Turin, Italy\\
$^{25}$ Dipartimento di Fisica e Astronomia dell'Universit\`{a} and Sezione INFN, Bologna, Italy\\
$^{26}$ Dipartimento di Fisica e Astronomia dell'Universit\`{a} and Sezione INFN, Catania, Italy\\
$^{27}$ Dipartimento di Fisica e Astronomia dell'Universit\`{a} and Sezione INFN, Padova, Italy\\
$^{28}$ Dipartimento di Fisica `E.R.~Caianiello' dell'Universit\`{a} and Gruppo Collegato INFN, Salerno, Italy\\
$^{29}$ Dipartimento DISAT del Politecnico and Sezione INFN, Turin, Italy\\
$^{30}$ Dipartimento di Scienze MIFT, Universit\`{a} di Messina, Messina, Italy\\
$^{31}$ Dipartimento Interateneo di Fisica `M.~Merlin' and Sezione INFN, Bari, Italy\\
$^{32}$ European Organization for Nuclear Research (CERN), Geneva, Switzerland\\
$^{33}$ Faculty of Electrical Engineering, Mechanical Engineering and Naval Architecture, University of Split, Split, Croatia\\
$^{34}$ Faculty of Engineering and Science, Western Norway University of Applied Sciences, Bergen, Norway\\
$^{35}$ Faculty of Nuclear Sciences and Physical Engineering, Czech Technical University in Prague, Prague, Czech Republic\\
$^{36}$ Faculty of Science, P.J.~\v{S}af\'{a}rik University, Ko\v{s}ice, Slovak Republic\\
$^{37}$ Frankfurt Institute for Advanced Studies, Johann Wolfgang Goethe-Universit\"{a}t Frankfurt, Frankfurt, Germany\\
$^{38}$ Fudan University, Shanghai, China\\
$^{39}$ Gangneung-Wonju National University, Gangneung, Republic of Korea\\
$^{40}$ Gauhati University, Department of Physics, Guwahati, India\\
$^{41}$ Helmholtz-Institut f\"{u}r Strahlen- und Kernphysik, Rheinische Friedrich-Wilhelms-Universit\"{a}t Bonn, Bonn, Germany\\
$^{42}$ Helsinki Institute of Physics (HIP), Helsinki, Finland\\
$^{43}$ High Energy Physics Group,  Universidad Aut\'{o}noma de Puebla, Puebla, Mexico\\
$^{44}$ Horia Hulubei National Institute of Physics and Nuclear Engineering, Bucharest, Romania\\
$^{45}$ Indian Institute of Technology Bombay (IIT), Mumbai, India\\
$^{46}$ Indian Institute of Technology Indore, Indore, India\\
$^{47}$ INFN, Laboratori Nazionali di Frascati, Frascati, Italy\\
$^{48}$ INFN, Sezione di Bari, Bari, Italy\\
$^{49}$ INFN, Sezione di Bologna, Bologna, Italy\\
$^{50}$ INFN, Sezione di Cagliari, Cagliari, Italy\\
$^{51}$ INFN, Sezione di Catania, Catania, Italy\\
$^{52}$ INFN, Sezione di Padova, Padova, Italy\\
$^{53}$ INFN, Sezione di Pavia, Pavia, Italy\\
$^{54}$ INFN, Sezione di Roma, Rome, Italy\\
$^{55}$ INFN, Sezione di Torino, Turin, Italy\\
$^{56}$ INFN, Sezione di Trieste, Trieste, Italy\\
$^{57}$ Inha University, Incheon, Republic of Korea\\
$^{58}$ Institute for Gravitational and Subatomic Physics (GRASP), Utrecht University/Nikhef, Utrecht, Netherlands\\
$^{59}$ Institute of Experimental Physics, Slovak Academy of Sciences, Ko\v{s}ice, Slovak Republic\\
$^{60}$ Institute of Physics, Homi Bhabha National Institute, Bhubaneswar, India\\
$^{61}$ Institute of Physics of the Czech Academy of Sciences, Prague, Czech Republic\\
$^{62}$ Institute of Space Science (ISS), Bucharest, Romania\\
$^{63}$ Institut f\"{u}r Kernphysik, Johann Wolfgang Goethe-Universit\"{a}t Frankfurt, Frankfurt, Germany\\
$^{64}$ Instituto de Ciencias Nucleares, Universidad Nacional Aut\'{o}noma de M\'{e}xico, Mexico City, Mexico\\
$^{65}$ Instituto de F\'{i}sica, Universidade Federal do Rio Grande do Sul (UFRGS), Porto Alegre, Brazil\\
$^{66}$ Instituto de F\'{\i}sica, Universidad Nacional Aut\'{o}noma de M\'{e}xico, Mexico City, Mexico\\
$^{67}$ iThemba LABS, National Research Foundation, Somerset West, South Africa\\
$^{68}$ Jeonbuk National University, Jeonju, Republic of Korea\\
$^{69}$ Johann-Wolfgang-Goethe Universit\"{a}t Frankfurt Institut f\"{u}r Informatik, Fachbereich Informatik und Mathematik, Frankfurt, Germany\\
$^{70}$ Korea Institute of Science and Technology Information, Daejeon, Republic of Korea\\
$^{71}$ KTO Karatay University, Konya, Turkey\\
$^{72}$ Laboratoire de Physique Subatomique et de Cosmologie, Universit\'{e} Grenoble-Alpes, CNRS-IN2P3, Grenoble, France\\
$^{73}$ Lawrence Berkeley National Laboratory, Berkeley, California, United States\\
$^{74}$ Lund University Department of Physics, Division of Particle Physics, Lund, Sweden\\
$^{75}$ Nagasaki Institute of Applied Science, Nagasaki, Japan\\
$^{76}$ Nara Women{'}s University (NWU), Nara, Japan\\
$^{77}$ National and Kapodistrian University of Athens, School of Science, Department of Physics , Athens, Greece\\
$^{78}$ National Centre for Nuclear Research, Warsaw, Poland\\
$^{79}$ National Institute of Science Education and Research, Homi Bhabha National Institute, Jatni, India\\
$^{80}$ National Nuclear Research Center, Baku, Azerbaijan\\
$^{81}$ National Research and Innovation Agency - BRIN, Jakarta, Indonesia\\
$^{82}$ Niels Bohr Institute, University of Copenhagen, Copenhagen, Denmark\\
$^{83}$ Nikhef, National institute for subatomic physics, Amsterdam, Netherlands\\
$^{84}$ Nuclear Physics Group, STFC Daresbury Laboratory, Daresbury, United Kingdom\\
$^{85}$ Nuclear Physics Institute of the Czech Academy of Sciences, Husinec-\v{R}e\v{z}, Czech Republic\\
$^{86}$ Oak Ridge National Laboratory, Oak Ridge, Tennessee, United States\\
$^{87}$ Ohio State University, Columbus, Ohio, United States\\
$^{88}$ Physics department, Faculty of science, University of Zagreb, Zagreb, Croatia\\
$^{89}$ Physics Department, Panjab University, Chandigarh, India\\
$^{90}$ Physics Department, University of Jammu, Jammu, India\\
$^{91}$ Physics Department, University of Rajasthan, Jaipur, India\\
$^{92}$ Physics Program and International Institute for Sustainability with Knotted Chiral Meta Matter (SKCM2), Hiroshima University, Hiroshima, Japan\\
$^{93}$ Physikalisches Institut, Eberhard-Karls-Universit\"{a}t T\"{u}bingen, T\"{u}bingen, Germany\\
$^{94}$ Physikalisches Institut, Ruprecht-Karls-Universit\"{a}t Heidelberg, Heidelberg, Germany\\
$^{95}$ Physik Department, Technische Universit\"{a}t M\"{u}nchen, Munich, Germany\\
$^{96}$ Politecnico di Bari and Sezione INFN, Bari, Italy\\
$^{97}$ Research Division and ExtreMe Matter Institute EMMI, GSI Helmholtzzentrum f\"ur Schwerionenforschung GmbH, Darmstadt, Germany\\
$^{98}$ Saha Institute of Nuclear Physics, Homi Bhabha National Institute, Kolkata, India\\
$^{99}$ School of Physics and Astronomy, University of Birmingham, Birmingham, United Kingdom\\
$^{100}$ Secci\'{o}n F\'{\i}sica, Departamento de Ciencias, Pontificia Universidad Cat\'{o}lica del Per\'{u}, Lima, Peru\\
$^{101}$ Stefan Meyer Institut f\"{u}r Subatomare Physik (SMI), Vienna, Austria\\
$^{102}$ SUBATECH, IMT Atlantique, Nantes Universit\'{e}, CNRS-IN2P3, Nantes, France\\
$^{103}$ Suranaree University of Technology, Nakhon Ratchasima, Thailand\\
$^{104}$ Technical University of Ko\v{s}ice, Ko\v{s}ice, Slovak Republic\\
$^{105}$ The Henryk Niewodniczanski Institute of Nuclear Physics, Polish Academy of Sciences, Cracow, Poland\\
$^{106}$ The University of Texas at Austin, Austin, Texas, United States\\
$^{107}$ Universidad Aut\'{o}noma de Sinaloa, Culiac\'{a}n, Mexico\\
$^{108}$ Universidade de S\~{a}o Paulo (USP), S\~{a}o Paulo, Brazil\\
$^{109}$ Universidade Estadual de Campinas (UNICAMP), Campinas, Brazil\\
$^{110}$ Universidade Federal do ABC, Santo Andre, Brazil\\
$^{111}$ University of Cape Town, Cape Town, South Africa\\
$^{112}$ University of Houston, Houston, Texas, United States\\
$^{113}$ University of Jyv\"{a}skyl\"{a}, Jyv\"{a}skyl\"{a}, Finland\\
$^{114}$ University of Kansas, Lawrence, Kansas, United States\\
$^{115}$ University of Liverpool, Liverpool, United Kingdom\\
$^{116}$ University of Science and Technology of China, Hefei, China\\
$^{117}$ University of South-Eastern Norway, Kongsberg, Norway\\
$^{118}$ University of Tennessee, Knoxville, Tennessee, United States\\
$^{119}$ University of the Witwatersrand, Johannesburg, South Africa\\
$^{120}$ University of Tokyo, Tokyo, Japan\\
$^{121}$ University of Tsukuba, Tsukuba, Japan\\
$^{122}$ University Politehnica of Bucharest, Bucharest, Romania\\
$^{123}$ Universit\'{e} Clermont Auvergne, CNRS/IN2P3, LPC, Clermont-Ferrand, France\\
$^{124}$ Universit\'{e} de Lyon, CNRS/IN2P3, Institut de Physique des 2 Infinis de Lyon, Lyon, France\\
$^{125}$ Universit\'{e} de Strasbourg, CNRS, IPHC UMR 7178, F-67000 Strasbourg, France, Strasbourg, France\\
$^{126}$ Universit\'{e} Paris-Saclay, Centre d'Etudes de Saclay (CEA), IRFU, D\'{e}partment de Physique Nucl\'{e}aire (DPhN), Saclay, France\\
$^{127}$ Universit\'{e}  Paris-Saclay, CNRS/IN2P3, IJCLab, Orsay, France\\
$^{128}$ Universit\`{a} degli Studi di Foggia, Foggia, Italy\\
$^{129}$ Universit\`{a} del Piemonte Orientale, Vercelli, Italy\\
$^{130}$ Universit\`{a} di Brescia, Brescia, Italy\\
$^{131}$ Variable Energy Cyclotron Centre, Homi Bhabha National Institute, Kolkata, India\\
$^{132}$ Warsaw University of Technology, Warsaw, Poland\\
$^{133}$ Wayne State University, Detroit, Michigan, United States\\
$^{134}$ Westf\"{a}lische Wilhelms-Universit\"{a}t M\"{u}nster, Institut f\"{u}r Kernphysik, M\"{u}nster, Germany\\
$^{135}$ Wigner Research Centre for Physics, Budapest, Hungary\\
$^{136}$ Yale University, New Haven, Connecticut, United States\\
$^{137}$ Yonsei University, Seoul, Republic of Korea\\
$^{138}$  Zentrum  f\"{u}r Technologie und Transfer (ZTT), Worms, Germany\\
$^{139}$ Affiliated with an institute covered by a cooperation agreement with CERN\\
$^{140}$ Affiliated with an international laboratory covered by a cooperation agreement with CERN.\\

\end{flushleft} 

\end{document}